\documentclass[sigconf,nonacm]{acmart}

\usepackage{setspace}
\usepackage{booktabs}
\usepackage{makecell}
\usepackage{graphicx}
\usepackage{subcaption}
\usepackage{multirow}
\usepackage[normalem]{ulem}
\useunder{\uline}{\ul}{}
\usepackage{algorithm}
\usepackage{algorithmicx}
\usepackage{algpseudocode}
\usepackage{amsmath}
\usepackage{enumitem}
\usepackage{float}
\usepackage{balance}

\usepackage{geometry}

\AtBeginDocument{%
  }

\begin{document}

\title{Sample Enrichment via Temporary Operations on Subsequences for Sequential Recommendation}


% Shu Chen
% chenshu20@email.szu.edu.cn
% College of Computer Science and Software Engineering, Shenzhen University

% Jinwei Luo
% luojinwei2016@email.szu.edu.cn
% College of Computer Science and Software Engineering, Shenzhen University
% Tencent Music Entertainment

% Weike Pan
% panweike@szu.edu.cn 
% College of Computer Science and Software Engineering, Shenzhen University

\author{Shu Chen}
\affiliation{%
  \institution{College of Computer Science and Software Engineering, Shenzhen University}
  % \institution{Shenzhen University}
  \city{Shenzhen}
  \country{China}
  }
\email{chenshu20@email.szu.edu.cn}

\author{Jinwei Luo}
\affiliation{%
  \institution{Shenzhen University}
  \institution{Tencent Music Entertainment}
  \city{Shenzhen}
  \country{China}
  }
\email{luojinwei2016@email.szu.edu.cn}

\author{Weike Pan}
\authornote{Corresponding author}
\affiliation{%
  \institution{College of Computer Science and Software Engineering, Shenzhen University}
  % \institution{Shenzhen University}
  \city{Shenzhen}
  \country{China}
  }
\email{panweike@szu.edu.cn}
% Jiangxing Yu
% jiangxingyu@tencent.com
% Tencent Music Entertainment

% Xin Huang
% kevinshuang@tencent.com
% Tencent Music Entertainment

% Zhong Ming
% mingz@szu.edu.cn
% College of Computer Science and Software Engineering, Shenzhen University
\author{Jiangxing Yu}
\affiliation{%
  \institution{Tencent Music Entertainment}
  \city{Shenzhen}
  \country{China}}
\email{jiangxingyu@tencent.com}

\author{Xin Huang}
\affiliation{%
  \institution{Tencent Music Entertainment}
  \city{Shenzhen}
  \country{China}}
\email{kevinshuang@tencent.com}

\author{Zhong Ming}
\affiliation{%
  \institution{Shenzhen University}
  \city{Shenzhen}
  \country{China}}
\email{mingz@szu.edu.cn}

\renewcommand{\shortauthors}{Chen et al.}

\begin{abstract}
\vspace{-0.5em}
  Sequential recommendation leverages interaction sequences to predict forthcoming user behaviors, crucial for crafting personalized recommendations. 
However, the true preferences of a user are inherently complex and high-dimensional, while the observed data is merely a simplified and low-dimensional projection of the rich preferences, which often leads to prevalent issues like data sparsity and inaccurate model training.
To learn true preferences from the sparse data, most existing works endeavor to introduce some extra information or design some ingenious models. 
Although they have shown to be effective, extra information usually increases the cost of data collection, and complex models may result in difficulty in deployment.
Innovatively, we avoid the use of extra information or alterations to the model; instead, we fill the \textbf{transformation space} between the observed data and the underlying preferences with randomness.
Specifically, we propose a novel model-agnostic and highly generic framework for sequential recommendation called \underline{s}ample \underline{e}nrichment via \underline{t}emporary \underline{o}perations on subsequences (SETO), which temporarily and separately enriches the transformation space via sequence enhancement operations with rationality constraints in training. 
The transformation space not only exists in the process from input samples to preferences but also in preferences to target samples.
We highlight our SETO's effectiveness and versatility over multiple representative and state-of-the-art sequential recommendation models (including six single-domain sequential models and two cross-domain sequential models) across multiple real-world datasets (including three single-domain datasets, three cross-domain datasets and a large-scale industry dataset).
\end{abstract}

\begin{CCSXML}
	<ccs2012>
	<concept>
	<concept_id>10002951.10003317.10003347.10003350</concept_id>
	<concept_desc>Information systems~Recommender systems</concept_desc>
	<concept_significance>500</concept_significance>
	</concept>
	</ccs2012>
\end{CCSXML}
\ccsdesc[500]{Information systems~Recommender systems}

\keywords{Sample Reconstruction, Sequential Recommendation, Sequence Augmentation, Cross-Domain Sequential Recommendation }


\maketitle

\vspace{-0.5em}
\section{Introduction}
\label{sec:introduction}

\begin{figure}[t]
    \centering
\includegraphics[width=0.94\linewidth]{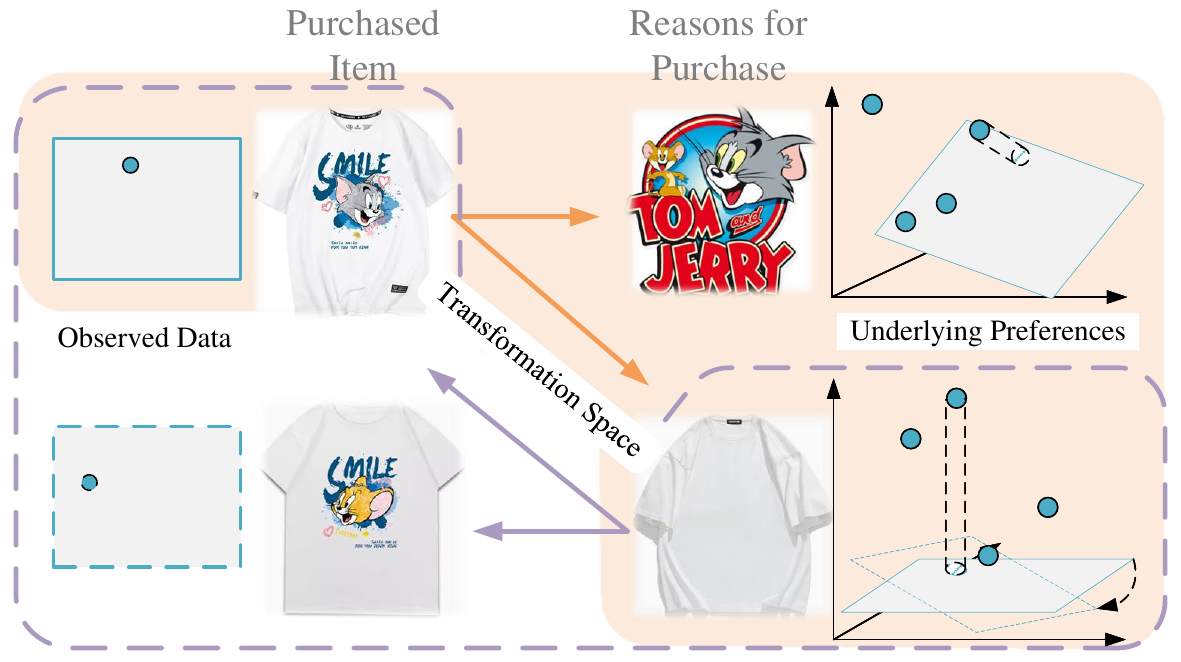}
    \centerline{(a) Observed data and underlying preferences.}
\includegraphics[width=1\linewidth]{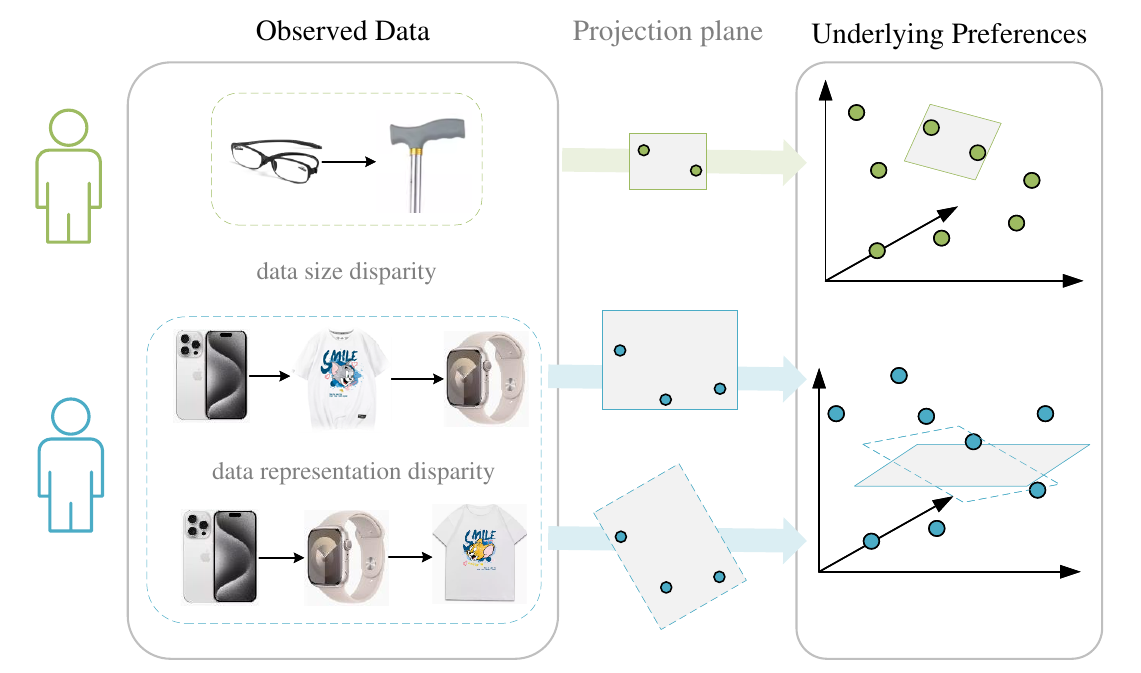}
    \centerline{(b) 2D planar points and 3D spatial points.}
    \vspace{-1.8em}
    \caption{An example that shows no one-to-one correspondence between the observed data and the underlying preferences but a certain transformation space (a).  Two-dimensional planar points represent the observed data and three-dimensional spatial points represent the true user preferences (b).}
    \vspace{-1em}
    \label{fig:example}
\end{figure}
\vspace{-0.4em}
Recommender systems are popular and vital tools of information retrieval and delivery, which endeavor to learn users' preferences from historical interaction data and then predict users' future interactions to offer a good recommendation.
Considering the sequential dependency between items, treating the historical interaction data as time-based sequences is a common practice, i.e., sequential recommendation (SR).
However, how to comprehensively extract valuable information from the observed data and how to improve the precision of preference learning with this information remain pivotal challenges.
Recent works utilize various deep neural networks to address these challenges, such as recurrent neural networks (RNNs)~\cite{GRU4Rec,LSTM, MV-RNN}, convolutional neural networks (CNNs)~\cite{Caser}, graph neural networks (GNNs)~\cite{SR-GNN,DGRec,CAGCN}, and Transformer~\cite{SASRec,FISSA}.

% Despite advancements in the field, data sparsity and incomplete modeling of the causes of user behavior remain significant challenges that limit the effectiveness of existing models.

Despite valid advances, models still face limitations in accurately translating data into preferences, particularly in scenarios with sparse data.
In a real-world example, as depicted in Figure~\ref{fig:example}(a), a model may speculate that a user purchasing a co-branded T-shirt does so due to an affinity for `Tom and Jerry', a preference for white T-shirts or other reasons. 
However, the model cannot be entirely certain of the true preferences due to the limited amount of data available for analysis.
This ambiguity often leads to inaccurate model training.
Essentially, small data can only establish a broad spectrum of preferences rather than pinpointing a precise value. 
In other words, the same data may correspond to different preferences. 
Similarly, the same preference may also correspond to different data.
Consider a scenario where a user expresses a desire to purchase a white T-shirt.  Numerous similar T-shirts could be chosen by him or her, but only one is shown as purchased in the final data, and this result is likely to be influenced by the contingencies (e.g., which item the user witnesses first). 
Obviously, there is not a direct one-to-one mapping between the observed data and the underlying preferences, leaving a certain \textbf{transformation space} between them.

\begin{figure}[t]
    \centering
    \includegraphics[width=1\linewidth]{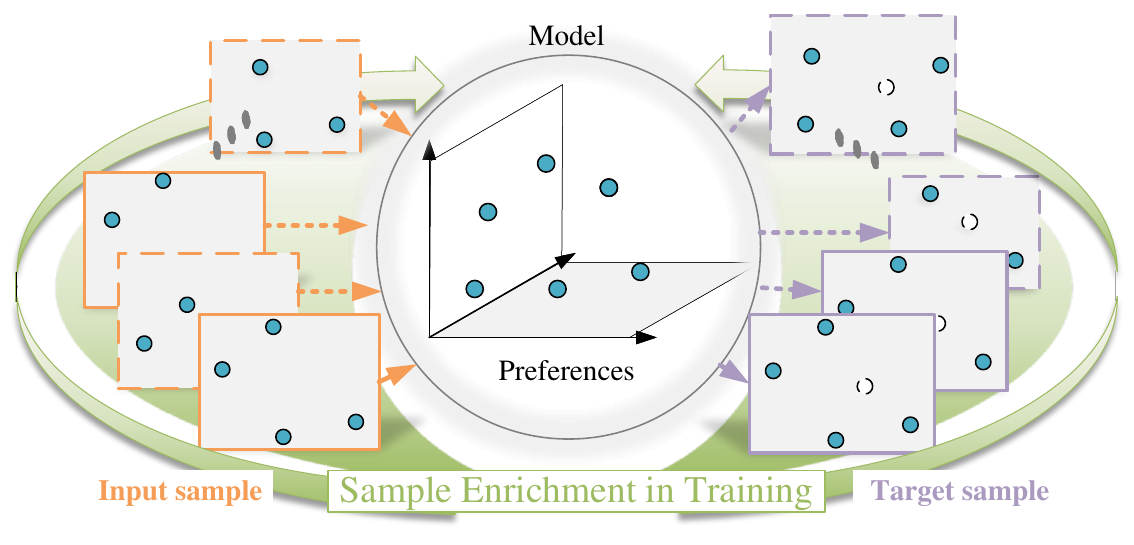}
\vspace{-2.5em}
    \caption{An illustration of sample enhancement in training, where a dashed box indicates a sample sequence that was altered (i.e., altering the position of items in the sequence or the length of the sequence), and a solid box indicates an original sample sequence.}
    \label{fig:sampler}    
\end{figure}

To a better understanding of this situation, we illustrate it from a multidimensional information perspective:
the true preferences of a user embody a kind of more complex and high-dimensional information, and the collected or observed data is only a low-dimensional projection onto a finite plane, as shown in Figure~\ref{fig:example}(b).
The duration and frequency of user interactions on the platform determine the size of the projection plane, thus determining the size of data we can collect or observe.
%The interaction data from an inactive user is so scarce that it represents only the tip of the iceberg of preferences.
% Furthermore, the orientation and position of the plane are not fixed and vary depending on the purchasing habits or contingencies. 
Furthermore, the difference in the orientation and position of the plane is influenced by the user personalization or contingencies, thus reflecting data representation disparity.
% For personalization, the projection planes of different users are with different orientations and positions in Figure~\ref{fig:example}(b).
% , and the same user also has different concerns about different kinds of goods.
For contingencies in the real world, the sequence of a user's behaviors may exhibit subtle fluctuations. 
For example, when presented with two items the user intends to purchase, the one encountered first is more likely to be bought first.
% So the orientation and position of the projection plane are needed to distinguish different users or different scenarios (domains).
The projection plane serves as a bridge between the visible data and the invisible preferences.
Based on this projection plane, we can metaphorically encapsulate the primary challenges inherent in sequential recommendation.
Too small a projection plane can cause data sparsity, and uncertainty in its orientation and position will result in inaccurate training.

Existing works improve recommendation performance mainly from two perspectives: one is to alleviate the data sparsity problem by introducing more information  (e.g., item attributes~\cite{S3Rec}, other domain's information~\cite{MGCL}, review~\cite{Review} etc.); the other is to design some ingenious models to get more accurate results~\cite{FISSA,FMLPRec}.
However, from a data perspective, the introduction of extra information not only places demands on the collection of data but also requires other components to help the model understand and consume it. 
From a model perspective, further improvement of the model often makes its structure more complex, which has a degree of difficulty in real deployment in production.
Considering that there is a \textbf{transformation space} between the observed data and the underlying preferences as mentioned above, we open up a novel perspective (i.e., neither introducing extra information nor changing the model) by proposing a model-agnostic and universal framework for sequential recommendation, i.e., \underline{s}ample \underline{e}nrichment via \underline{t}emporary \underline{o}perations on subsequences (SETO, in short).
Our SETO utilizes the transformation space in the two steps, including (i) from input samples to preferences and (ii) from preferences to target samples, to temporarily perform proper enhancement operations on the partitioned input and target subsequences (instead of the original sequence) in each iterative training, as shown in Figure~\ref{fig:sampler}.
Moreover, based on the effect of the position and size of the projection plane on the observed data, we propose two operations with rationality constraints using probabilistic and random selection, i.e., $Swap$ and $Removal$, and apply them to the input and target subsequences. 
To validate the effectiveness and versatility of our SETO, we apply it to several state-of-the-art sequential recommendation models (including six single-domain sequential models and two cross-domain sequential models) and then conduct extensive experiments on six real-world datasets (including three single-domain datasets and three cross-domain datasets), as well as a large-scale industry dataset.
We summarize our main contributions as follows: 
\begin{itemize}[leftmargin=1em]
    \item We propose a novel model-agnostic framework SETO, which leverages the transformation space between the observed data and the true user preferences, enriching training samples in iterative training via temporary data enhancement operations.
    \item We follow the assumptions about the effect of the position and size of the projection plane on the observed data and design two sequence augmentation operations with rationality constraints, i.e., $Swap$ and $Removal$, which operate on the partitioned subsequences rather than the original sequence.
    \item Through extensive empirical studies, we demonstrate the effectiveness and versatility of our SETO over multiple representative and state-of-the-art sequential recommendation models (including six single-domain sequential models and two cross-domain sequential models) across multiple datasets (including three single-domain datasets, three cross-domain datasets and a large-scale industry dataset).
\end{itemize}

% \noindent \textbf{(1)} We propose a novel model-agnostic framework SETO, which leverages the transformation space between the observed data and the true user preferences, enriching training samples in iterative training via temporary data enhancement operations.

% \noindent \textbf{(2)} We follow the assumptions about the effect of the position and size of the projection plane on the observed data and design two sequence augmentation operations with rationality constraints, i.e., $Swap$ and $Removal$, which operate on the partitioned subsequences rather than the original sequence.

%  \noindent \textbf{(3)} Through extensive empirical studies, we demonstrate the effectiveness and versatility of our SETO over multiple representative and state-of-the-art sequential recommendation models (including six single-domain sequential models and two cross-domain sequential models) across multiple datasets (including three single-domain datasets, three cross-domain datasets and a large-scale industry dataset).

\vspace{-1.5em}
\section{Related Work}
In this section, we overview the related work from two perspectives: sequential recommendation and sequence augmentation for sequential recommendation.

\subsection{Sequential Recommendation}
Before deep learning became widely applied in recommender systems, sequential recommendation models primarily relied on Markov chains (MCs)~\cite{using} or matrix factorization (MF)~\cite{BPR} to capture item transitions and identify users' short-term or long-term interests.
To better model the interaction relationship between users and items, translation-based models~\cite{TransRec, CTransRec} are proposed. 
Among deep learning networks, recurrent neural networks (RNNs) are first employed to incorporate the temporal sequential information of a sequence ~\cite{GRU4Rec,LSTM,MV-RNN}. 
% Due to the characteristic of vanishing gradients in the RNN-based models, 
Then convolutional neural networks (CNNs) are utilized for sequential recommendation~\cite{Caser}. 
Since the advent of Transformers~\cite{Transformer}, several models leveraging Transformer architecture~\cite{SASRec, FISSA, AdaMCT} achieve significant performance improvements. 
For instance, SASRec~\cite{SASRec} is a primary and representative model, which gets much attention from recent works~\cite{PinnerFormer,meta,zhou2024contrastive}.
Additionally, some recent works~\cite{SR-GNN,DGRec,CAGCN,LTGNN} address sequential recommendation via graph neural networks (GNNs), which involves constructing a directed graph where each user interaction serves as a node. 
However, GNN-based methods are proven to perform better on short sequences, so they are often applied to the session-based recommendation~\cite{session-based-survey}.

To address the data sparsity issue, recent works attempt to introduce extra information to improve the accuracy of recommendations, in which cross-domain information garnered a lot of attention, thus forming a new topic called cross-domain sequential recommendation (CDSR)~\cite{CDSRsurvey}.
Most models~\cite{pi-Net,CD-SASRec,MGCL,DREAM,TJAPL,AMID} mix all sequences from different domains in chronological order and further learn the sequential relationships between items of different domains.

Our proposed SETO is a model-agnostic approach that aims to improve the performance of a certain sequential recommendation method,  without requiring any modifications to the model's structure or loss function.

\begin{table}[t]
  \centering
  \caption{A brief review of the works on sequence augmentation for sequential recommendation from the perspectives of model agnostic and model dependent, as well as whether using the context information.}
   \vspace{-0.7em}
\label{tab:Classification of related works}
    % \vspace{-0.4em}
    \resizebox{1\linewidth}{!}{
    \begin{tabular}{lcc}
    \toprule
          & w/o context & w/ context \\
    \midrule
    Model agnostic  & RSS~\cite{RSS}, {\bf SETO (proposed)} & $\backslash$ \\
    \midrule
    Model dependent & DuoRec~\cite{DuoRec}, DiffuASR~\cite{DiffuASR} , etc. & S$^3$Rec~\cite{S3Rec}, etc. \\
    \bottomrule
    \end{tabular}
    }
  \label{tab:addlabel}%
\end{table}%

% \vspace{-0.5em}
\subsection{Sequence Augmentation for Sequential Recommendation}
% sequence augmentation is a popular technique in several fields (e.g., imaging, natural language processing, etc.).
In Table \ref{tab:Classification of related works},  we summarize the related works of sequence augmentation for sequential recommendation from two perspectives: whether the augmentation is only at the data level and not related to the model, and whether it involves context beyond the user and interacted item sequences. 
% \textcolor{blue}{
Most of the recent works exploit the augmented sequences via contrastive learning (CL)~\cite{DuoRec, CL4SRec,TiCoSeRec}, which usually designs an additional loss function in the models. 
% to pull in the prediction results of the similarly generated sequences or enable the different sequences to deviate from. 
However, \cite{zhou2024contrastive} shows that using a sliding window alone could outperform CL and most of the other enhancement strategies are inferior to CL. 
Notice that since the sliding window can be viewed as an additional operation on other operations (e.g., cut, insert, etc.)~\cite{Caser}, we do not use it in order to study the effectiveness and universality of our SETO.
% }
% CL-based methods have been shown to be effective, but the best performance could also be achieved by sequence enhancement methods alone, which would benefit from less training time and less computational cost~\cite{zhou2024contrastive}.
Moreover, some models, e.g., DiffuASR~\cite{DiffuASR}, pre-train a new model to generate a new sequence, thus achieving augmentation.
% Our SETO is distinguished from those models, which operate at the model level, while our approach only modifies the data level.

Additionally, some models enhance the sequences by utilizing some additional information, such as the CL-based TiCoSeRec~\cite{TiCoSeRec}, which uses time intervals to transform non-uniformly distributed sequences into uniformly distributed time sequences, and S$^3$Rec~\cite{S3Rec} that utilizes the attributes of the items. 
In fact, cross-domain information is also one type of additional information, but in this paper, we treat it as an extended scenario for our SETO's application since only the ID information of the interacted items is used.

In contrast to the models mentioned above, our SETO is model-agnostic and only utilizes the most basic information (i.e., the interaction sequences of users). 
We find only the RSS~\cite{RSS} approach resembles our work, which randomly selects any item of the sequence as the target item and not just the last item.  
However, RSS improves the backbone model only for a limited time (i.e., 1 hour) and does not guarantee that it will remain valid when the model is fully converged.
Our SETO fully considers the hidden transformation space in the two steps of data-to-preference and preference-to-data, and disposes of sequence augmentation operations with rationality constraints to enhance the effect of the state-of-the-art sequential recommendation models.
% \textcolor{blue}{
For more differences between our SETO and the traditional enhancement strategies, please see the Appendix~\ref{sec:vs}.
\begin{figure}
    \centering
    \includegraphics[width=1\linewidth]{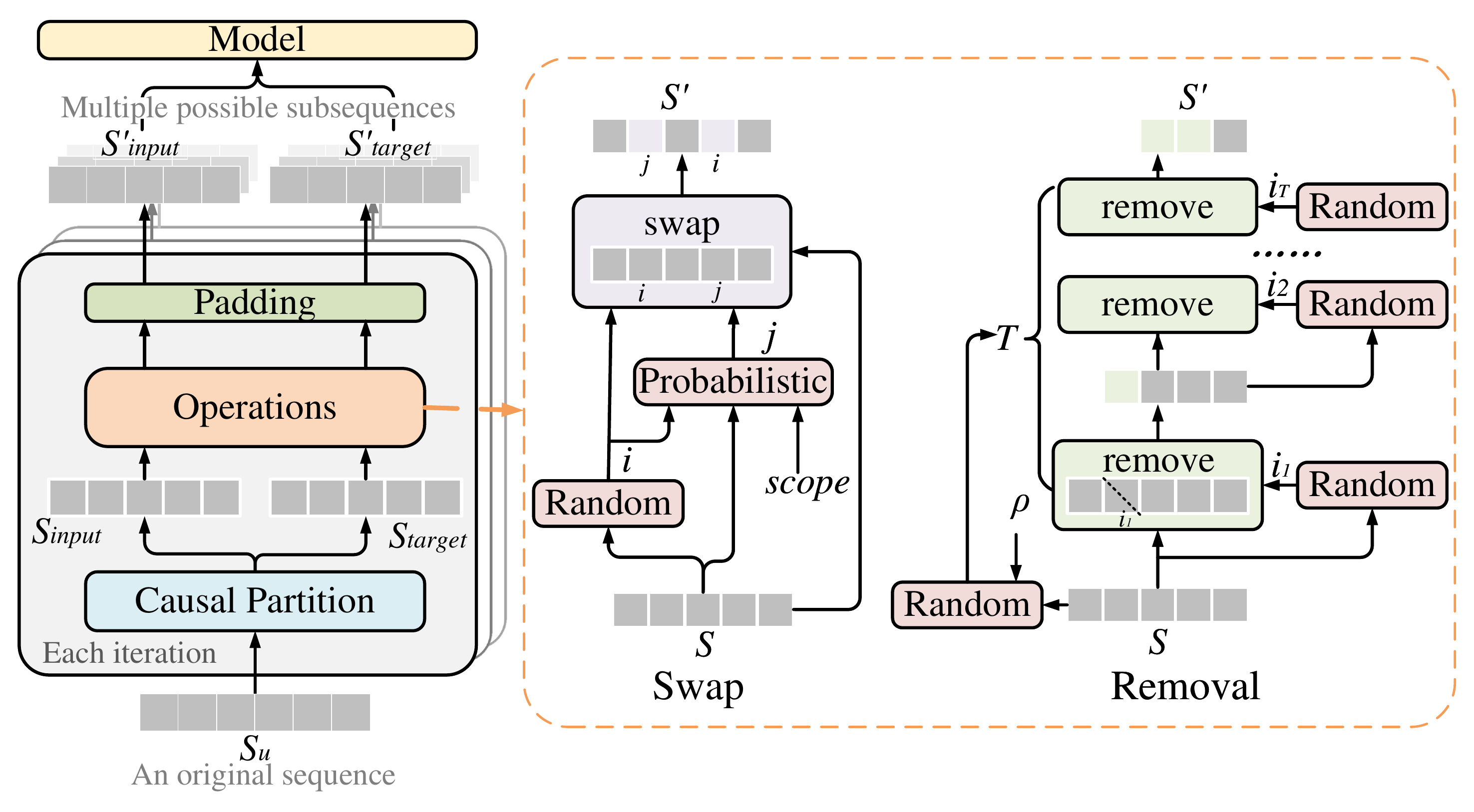}
    \vspace{-1.7em}
    \caption{The overview of our SETO on the left, and two operations illustrated on the right.  
    $S'$ is the new sequence after those operations are applied on $S$. We do the temporary operations with rationality constraints on the original input and target subsequences in each training iteration, thus enriching diverse training samples. 
    % Our SETO combines random and multiple iterative training to enrich diverse training samples.
    }
    \label{fig:overview}
\end{figure}
\section{Methodology}
A typical recommender system consists of two sets: users $U$ and items $I$. Each user has a sequence of interacted items, denoted $ S_u=\{s_1,s_2,\cdots,s_{|S_u|}\} $.  
In cross-domain sequential recommendation (CDSR), the user set usually consists of overlapping users who interact with items in all domains, and the interaction sequences in two different domains can be denoted as $ S_u^X=\{s_1^X,s_2^X,\cdots,s_{|S_u^X|}^X\} $ and $ S_u^Y=\{s_1^Y,s_2^Y,\cdots,s_{|S_u^Y|}^Y\} $, where $X$ and $Y$ represent the target and source domains, respectively.
Based on the above information, we aim to predict the item that a user will interact with at the next step.
% In deep learning models, the construction of reasonable and effective sequence samples often greatly affects the final recommendation performance. 
In order to address the challenges explained in Section \ref{sec:introduction}, we propose a novel and highly generic framework for sequential recommendation called \underline{s}ample \underline{e}nrichment via \underline{t}emporary \underline{o}perations on subsequences (SETO). It utilizes the transformation space between the observed data and the underlying preferences to temporarily perform enhancement operations with rationality constraints on the partitioned input and target subsequences in each iterative training, thus optimizing the model training process.

In Section \ref{subsec:Processing Procedure}, we provide a comprehensive explanation of how samples are enriched in our SETO and provide an overview in Figure \ref{fig:overview}. 
We also the pseudocode of our SETO in Appendix~\ref{sec:algorithm} in the auxiliary material.

\subsection{Processing Procedure}
\label{subsec:Processing Procedure}
\subsubsection{\textbf{Sequential Causal Partitioning}}
\label{sec:SequentialCausalPartitioning}
Training samples typically consist of three components: the input of a model, positive samples (also known as target output), and negative samples. In this paper, we focus on the input part and the target output part to reconstruct the training samples.
In the early stage of sequential recommendation research, most studies~\cite{FPMC, Fossil} divide the items of an original user sequence into two non-overlapped parts, i.e., treat the last item as the target subsequence and all previous items as the input subsequence. 
With the advent of the sequence-to-sequence (Seq2Seq) paradigm, some models~\cite{SASRec,FISSA,BERT4Rec} (especially those based on Transformer) began to apply the cross-based construction method, which enables all items except for the last one as the input subsequence, and lets all items except for the first one as the target subsequence. 
In order to maximize the retention of item information and achieve more space to operate, we follow the cross-based training sample construction method as our causal partition method.

% We utilize the cross-based method in the causal partitioning of our SETO, which can provide more training opportunities than the separation-based method because of has overlapped parts and is also more semantically relevant to predicting the next item than the mask-based method.
% It maximizes items' information retention in the two subsequences after partition while maintaining the transitive relationship between the previous item and the next item.
For a user $u$'s sequence $ S_u=\{s_1,s_2,\cdots,s_{|S_u|}\} $, we divide it into two parts: $ S_{input}=\{ s_1,s_2,\cdots,s_{|S_u|-1}\} $ and $ S_{target}=$\{$ s_2,$ $s_3,$$\cdots,$ $s_{|S_u|}$ $\}$. 
 $S_{input}$ consists of all items in $S_u$ except for the last one, while $S_{target}$ comprises all items in $S_u$ except for the first one. We use $S_{input}$ as the input during training and aim to generate $S_{target}$ as its positive output. For the standardization of sequential model operations, we define a maximum sequence length of $L$. In case the length of $S_{u}$ is longer than $L$, we follow most previous practices~\cite{SASRec, FISSA} to truncate them to length $L$. 
% However, if the length of $S_{u}$ is already less than $L$, we temporarily unalter them. 
Note that we use $S$ to represent an ordinary sequence and $CP(\cdot)$to represent the causal partition of $S$  as follows,
\begin{equation}
  S_{input},S_{target}=CP(S_{|S|-L+1:|S|}),
\end{equation}
where $|S|-L+1:|S|$ denotes that we take only the last $L$ items of the original sequence. And if the length of the original sequence is less than $L$, we directly utilize the entire original sequence $S$.

\subsubsection{\textbf{Two Operations on Subsequences}}
\label{subsec:TwoAtomicOperations}
Based on the assumption (i.e., data can be viewed as a projection of a user's true preferences on a projection plane) mentioned in Section~\ref{sec:introduction}, we design two sequence augmentation operations (i.e., $Swap$ and $Removal$), corresponding to fluctuations in the position and size of the projection plane, respectively.
To ensure the plurality and rationality of the new samples enriched, those operations contain both random and probabilistic choices. 

 In order to realize the probabilistic choice, we first define $f(n,i)$ as the probability function inspired by \cite{RSS}. This function returns a probability value, where $i$ denotes the relative position of the item within a total length of $n$. The smaller the value of $i$, the higher the probability value obtained in a single choice. The probability values obtained for all positions are increasing or decreasing, with a summation of 1.  Numerous functions can satisfy this property, and an example of which is provided below,
\begin{equation}
\label{equ:probabilityFunction}
    f(n,i)=\frac{\alpha^i}{\sum_{j=0}^{n} \alpha^j}
\end{equation}
where we use the power operation to implement the increasing or decreasing characteristic, and $ \alpha \in [0,1]$ is the parameter that decides the probability of selecting the nearer items. 

Next, we introduce the $Swap$ and $Removal$ operations, where we consider $S$ as an ordinary to-be-processed sequence, $n$ is its length and $S'$ denotes the new sample sequence constructed by those operations.

\noindent\textbf{Swap.}
In fact, the order relationship between similar items that a user interacts with in a short period of time is often not strict. For example, when a user compares two different brands of cell phones,  both have the possibility of being clicked first.
Hence, by swapping the positions of nearby items,  we can get the new sample to fill the transformation space between the observed data and the underlying preferences.
Note that the tolerance in an item sequence in typical recommender systems is often much higher than that in a word sequence in NLP due to the language habits and grammar rules.
For a sequence $S=$$\{s_1,$$s_2,$$\cdots,$$s_i,$$\cdots,$$s_j,$$\cdots,$$s_n\}$, we randomly select a pivot item $s_i$ and exchange its position with another nearby item $s_j$ that is selected by the probability function in Equation~\ref{equ:probabilityFunction}, and then we get the new disordered sequence $S'=\{s_1,s_2,\cdots,s_j,\cdots,s_i,\cdots,s_n\}$. 
 Additionally, for the selection of the position of the second item $s_j$ in the above operation, we set a parameter $scope \in [0,1]$, which indicates the percentage of the entire sequence length, used to limit the farthest position of the two exchanged items. Note that a random space is still reserved when selecting items within the range.  
 We use $Random(:)$ and $Prob(:,f)$ to denote random and probabilistic choices, respectively, in $S$ as follows, 
\begin{gather}
    i=Random(1:n)\\
    j=Prob(i-\lfloor n \times scope\rfloor:i+\lfloor n\times scope\rfloor,f(\lfloor n \times scope\rfloor, |i-j|))
\end{gather}    
where the colon indicates the range of choices, and $f(\cdot)$ denotes the probability of picking a position from Equation~\ref{equ:probabilityFunction}.
% where the farthest range of the exchange position is indicated by the square brackets. If $scope=1$, it means that $s_j$ can be any item in the whole sequence. If $scope=0$, $s_i$ can only exchange with itself without any changes. 
The parameter $scope$ ensures a certain degree of rationality for our exchange operations. For example, it would be unreasonable to exchange an item purchased recently with one purchased some years ago. And by setting it in the form of a percentage rather than a fixed value, it can be applied to sequences with different lengths.

\noindent\textbf{Removal.}
Similarly, we believe that the subsequence after taking a small amount of removal operations could equally serve as an interaction sequence for a user.
For a sequence $S=\{s_1,s_2,\cdots,s_{n}\}$, we can remove some items temporarily and randomly in a given iteration, thus obtaining new sample sequences.
In each new iteration, we randomly choose the number of items to be removed and which ones to remove. 
For the number of items to be removed, we utilize a parameter $\rho$ to specify the maximum percentage of censored items relative to the total number of items in the sequence.
For example,  if $\rho \times n=3$, then we can remove $k$ ($0\leq k \leq 3$) items in the sequence by random selection. The $Removal$ operation allows the model to better learn the interrelationships between two items at different intervals, which thus keeps the learning of the item from being limited to a fixed length of sequence.
Note that if an item is removed,  all items before the removed item are moved back, with no spaces or padding. 
% For example, if we decide to remove two items, and $\{s_i, s_j\}$ were randomly selected, the sequence after removal is $S'=\{s_1,\cdots,s_{i-1},s_{i+1},\cdots,s_{j-1},s_{j+1},\cdots,s_{n}\}$, no spaces or padding.
% Sequential models~\cite{SASRec,FISSA} learn the position embeddings of items based on their relative position in the sequence, not on the absolute time. 
% we keep the learning of the item from being limited to a fixed length of sequence, considering the immobility of realistic interactions.
We take multiple removal operations to the original sequence $S$ based on a randomly obtained number $T$ as follows,
\begin{gather}
    T=Random(0:\lfloor \rho \times n \rfloor)\\
    S^{(t)}=remove^{(t)}(S^{(t-1)},i_t),t \in \{1,2,\cdots,T\}\\
    i_t=Random(1:|S^{(t-1)}|)
\end{gather}
where $S^{(t)}$ is the sequence after removing the $t$-th item, and the $i_t$-th item is selected for removal in $S^{(t-1)}$. If $T=0$, $S'=S$.

Note that for either $Swap$ or $Removal$, we design with the possibility of preserving the original sequence unaltered. The results of the two operations are only valid for the current iteration, and the sequence reverts to its original form before the next iteration.
As a result, we can take full advantage of the diverse samples enriched in the iterations. And in case the altered parts contain some noise, they will not affect the modeling training much as they are limited in one single iteration.

\subsubsection{\textbf{Padding Short Sequences}}
As above, we do not handle sequences shorter than $L$. Additionally, in the $Removal$ method, the length of the processed sequences also decreases. To maintain consistency during model training, we pad the sequences at the beginning to reach a maximum length of $L$ as follows,
\begin{equation}
\begin{split}
        &S=\{s_1,s_2,\cdots,s_{|S|}\}\\
        \rightarrow &S'=\{<pad>,\cdots,<pad>,s_1,\cdots,s_{|S|}\}
\end{split}
% S=\{s_1,s_2,\cdots,s_{|S|}\}
%         \rightarrow S'=\{<pad>,\cdots,<pad>,s_1,\cdots,s_{|S|}\}
\end{equation}
where the length of the processed sequence is $L$, i.e., $|S'|=L$. Finally, the processed input subsequences and target subsequences can be used as the training samples for model training. 
% We provide an overview in Figure \ref{fig:overview} to illustrate the detailed process of our SETO.

\subsection{Discussions}
\label{subsec:Discussion}

% \begin{figure*}[t]
%     \centering
%     \includegraphics[width=0.58\textwidth]{fig/sampleNum.pdf}
%     \caption{Statistics on the change in the number of sample types in an iterative run. We use a user's sequence of length 11 in the real dataset (Foursquare) as an example. Owing to randomness in operations, we run it 5 times and take the average. ``ori" denotes that only the original sequence is processed, ``input'' and ``target'' respectively denote that the input and output sequences are processed separately, while ``all'' denotes that both sequences of the causal split are processed.}
%     \label{fig:sampleNum}
% \end{figure*}

% In this section, we focus on analyzing the advantages of the two operations and the differences in the number of kinds of samples that can be constructed from different operations working on different objects in iterative training.  
To verify the reliability of enriched samples in the transformed space between data and preferences, we design two sequence augmentation operations, i.e., $Swap$ and $Removal$, based on the location and size of the projection plane.
In this section, we focus on analyzing the advantages and the different characteristics of the two operations,  which temporarily enrich the sample sequences at each iteration.  

Comparing the two operations, the $Removal$ operation provides an advantage over the $Swap$ operation as it can preserve the relative position of the items in a single direction, making it more suitable for datasets with strong sequential characteristics, such as those containing geographical location information (e.g., the performance of $Removal$ on Foursquare in Table~\ref{tab:SETO_baselines}). 
In contrast to direct removal from the original sequence, removal of an item from a subsequence does not cause the item to completely disappear from the entire training sample, which in turn achieves diversified construction while preserving the information of the items in the sequence.
On the other hand, the $Swap$ operation does not lose any items and can be more effective when a majority of items in a user sequence are significant to user preferences.
Moreover, $Swap$ can construct more sample types than $Removal$ because it does not reduce the number of items in the sequence. Operating separately on both the input and target subsequences after causal splitting can result in more training samples than only handling the original sequence as often adopted in previous related works.

% We take an example of the way training samples are handled in SASRec~\cite{SASRec}, i.e., predict the second item according to the first item, predict the third item according to the first two items, and extrapolate to predict the last item based on the entire input subsequence. 
% We treat a different predictive recursive relationship as a sample type, and then count the change in the number of sample types as different operations are implemented on different objects with running iterations in the Figure~\ref{fig:sampleNum}.
% Overall, $Swap$ can construct more sample types than $Removal$ because it does not reduce the number of items in the sequence. Operating separately on both input and target subsequences after causal splitting can result in more training samples than only handling the original sequence.

\begin{table}
 \caption{The statistics of the datasets.}
 \vspace{-1em}
  \label{tab:datasets}
 \centering
%  \resizebox{1\linewidth}{!}{
% \begin{tabular}{ccccccc}
% \toprule
% Dataset    &\#Users & \#Items  & \#Interactions & \begin{tabular}[c]{@{}c@{}}Avg.\\ length\end{tabular} &\begin{tabular}[c]{@{}c@{}} Max.\\length\end{tabular} & Density\\ \midrule
% Foursquare & 22,748 & 11,146 & 145,106      & 6.38  &25                                                    & 0.06\%  \\
% Games      & 29,341 & 23,464 & 280,945      & 9.58    &860                                                  & 0.04\%  \\
% Beauty     & 40,226 & 54,542 & 353,962      & 8.80      &  293                                              & 0.02\%  \\  \bottomrule
% \end{tabular}}
\resizebox{1\linewidth}{!}{
\begin{tabular}{ccccccc}
\toprule
 & Dataset & \#Users & \#Items & \#Interactions & \begin{tabular}[c]{@{}c@{}}Avg.\\Length\end{tabular} & Density \\ \midrule
 &  Foursquare & 22748 & 11146 & 145106 & 6.38 & 0.06\% \\ 
 & Games & 29341 & 23464 & 280945 & 9.58 & 0.04\% \\
\multirow{-3}{*}{\begin{tabular}[c]{@{}c@{}}Single\\ Domain\end{tabular}} &  Beauty & 40226 & 54542 & 353962 & 8.08 & 0.02\% \\ \midrule
 % & Dataset &  \begin{tabular}[c]{@{}c@{}}\#Overlapped\\Users\end{tabular} & \#Items & \#Interactions & \begin{tabular}[c]{@{}c@{}}Avg.\\Length\end{tabular} & Density \\ \midrule
 & Movie &  & 60902 & 462314 & 42.30 & 0.07\% \\
 & CD &  & 94171 & 348746 & 31.91 & 0.03\% \\
\multirow{-3}{*}{\begin{tabular}[c]{@{}c@{}}Cross\\ Domain\end{tabular}} & Book & \multirow{-3}{*}{\begin{tabular}[c]{@{}c@{}}10929\\(Overlapped)\end{tabular}} & 242363 & 615912 & 56.36 & 0.02\% \\
\bottomrule
\end{tabular}}
\end{table}

\section{Experimental Setup}
In this section, we provide a detailed description of our experimental setup, which aims to better answer the following questions:
 \label{sec:rq}
\begin{enumerate} 
    \item[\textbf{RQ1}]\label{RQ1} Can our SETO improve the performance of the representative and state-of-the-art models?
     \item[\textbf{RQ2}]\label{RQ2}   Can our SETO be directly applied to cross-domain sequential recommendation?
    
    \item[\textbf{RQ3}]\label{RQ3} Does our SETO also perform well on a large-scale industry dataset?
    \item[\textbf{RQ4}]\label{RQ4} 
    % Are our two atomic operations valid compared to complete randomization? 
    What is the performance of our SETO when only processing the input subsequences or the target subsequences?
    \item[\textbf{RQ5}]\label{RQ5} What impact do variations in $scope$ of the $Swap$ method and variations in $\rho$ of the $Removal$ method have on recommendation performance? 
\end{enumerate}
We also experimentally analyze the training efficiency of our SETO and some related works (i.e., RSS~\cite{RSS} and DuoRec~\cite{DuoRec}) on backbone model SASRec~\cite{SASRec}.
Please see Appendix~\ref{sec:efficiency} in the auxiliary material. 
% \vspace{-0.5em}
\subsection{Datasets}
To demonstrate our SETO's effectiveness in improving the performance of a backbone model,
we use three datasets in real-world scenarios, i.e., Foursquare, Games and Beauty. 
To further validate the universality of our SETO, we apply it to two cross-domain sequential recommendation models on three corresponding datasets, i.e., Movies, CD, and Books.

\textbf{Foursquare}\footnote{\href{url}{https://archive.org/details/201309\_foursquare\_dataset\_umn}} is a dataset of the eponymous social application, which records the check-in data of different users in different places. 
% We utilize its historical data to predict the next place where each user will check in. 
Note that this dataset exhibits stronger sequential nature compared with other e-commerce datasets, due to the inclusion of the geographical location information. 
% Therefore, it is more suitable for the processing method, i.e., $Removal$, that can preserve the relative order.
\textbf{Amazon}\footnote{\href{url}{https://cseweb.ucsd.edu/\url{~}jmcauley/datasets.html\#amazon\_reviews}} is a well-known e-commerce platform, which categorizes goods according to their categories.
We select \textbf{Games} and \textbf{Beauty}, which exhibit high sparsity and are frequently used in previous works~\cite{SASRec, FISSA, BERT4Rec}. 
% Specifically,  we consider the reviewed items by a user and predict the next item that the user will examine.
Additionally, \textbf{Movies}, \textbf{CD} and \textbf{Books} are treated as the cross-domain datasets in Section~\ref{sec:Q4}, which are very popular with state-of-the-art models~\cite{MGCL,CD-SASRec} in CDSR.
For the single-domain datasets, we adopt the processed data\footnote{\href{url}{https://csse.szu.edu.cn/staff/panwk/publications/FISSA/}} from FISSA~\cite{FISSA}. 
For the cross-domain datasets, we follow  MGCL~\cite{MGCL} in data processing\footnote{\href{url}{https://csse.szu.edu.cn/staff/panwk/publications/MGCL/}}.
Additional details about the datasets are provided in Table \ref{tab:datasets}.

\vspace{-0.6em}
\subsection{Evaluation Metrics}
% To provide a more intuitive demonstration of the effectiveness of our SETO, we have selected two evaluation measures as follows:
We choose two evaluation metrics to study the effectiveness of our SETO.
(i)~Recall@10: It represents the hit rate in the top 10 items but does not consider the correct position of the item among these 10 items.
(ii)~NDCG@10: 
It is the abbreviation for normalized discounted cumulative gain at 10, which considers the relative positions of the correct items among the top 10 items. 
% The score increases as the target item is ranked higher.
Note that the evaluation measure of RSS~\cite{RSS} (c.f. 1-hour limited training) does not fully demonstrate the training performance of the model. Hence,  we adopt the above two commonly used evaluation metrics.
% Note that following FISSA~\cite{FISSA}, we randomly sampled 100 items that the user has not interacted with as the partial candidate set for each user.  Moreover, for a more comprehensive evaluation, we conduct additional experiments on the complete item candidate set in Section \ref{subsec:entire candidate}. 
% For each dataset, we adopt a leave-one-out evaluation methodology, where the second last item is used as validation and the last item is held for testing.

\vspace{-0.6em}
\subsection{Baselines}
 Due to the non-intrusive and highly generic feature of our SETO, we combine it with six representative and state-of-the-art models in single-domain sequential recommendation based on different techniques to demonstrate its effectiveness: \textbf{Caser}~\cite{Caser}, \textbf{SR-GNN}~\cite{SR-GNN}, \textbf{SASRec}~\cite{SASRec}, \textbf{FISSA}~\cite{FISSA}, \textbf{FMLP-Rec}~\cite{FMLPRec} and \textbf{DiffuRec}~\cite{DiffuRec}. 
To further verify the effectiveness and versatility, we adopt two state-of-the-art CDSR models, i.e., \textbf{MGCL} \cite{MGCL} and \textbf{TJAPL}~\cite{TJAPL}. 
 For detailed descriptions of the backbone models, please see Appendix~\ref{sec:backbone} in the auxiliary material.
 
 In addition, we include \textbf{RSS}~\cite{RSS} in our experiments as a closely related method, which proposes using the beginning of a sequence as a learning target.
Though model-dependent sequence augmentation is distinguished from our non-intrusive frameworks can be found in Table~\ref{tab:Classification of related works}, we still include the state-of-the-art model, i.e., \textbf{DuoRec}~\cite{DuoRec}, as a reference. 
Specifically, it proposes a model-dependent augmentation based on dropout and a novel sampling strategy, where sequences having the same target item are chosen as hard positive samples. 
Note that for a fair comparison, we set the training sample construction method of DuoRec to the same cross-based method as ours.

% Note that although  TiCoSeRec~\cite{TiCoSeRec} is a recent model that designs five operators for sequence augmentation, we do not consider it as our baseline. This is because it utilizes the time interval information, which is not available in our studied problem.
% As for the representative model of the mask-based construction method, i.e., BERT4Rec~\cite{BERT4Rec}, a related work~\cite{StudyofBERT4Rec} has shown that it needs a very long time to achieve good performance, while its prediction performance is inferior to SASRec~\cite{SASRec} without sufficient training.
% For the other traditional sequential recommendation models, such as the MF-based models (i.e., BPRFM~\cite{BPR}, FISM~\cite{FISM}, FPMC~\cite{FPMC}and Fossil~\cite{Fossil}) and the DL-based models (i.e., GRU4Rec+~\cite{GRU4Rec}, Caser~\cite{Caser} and CAR~\cite{CAR}), we copy the results of them from \cite{FISSA} because we use the same datasets and evaluation protocol.

\vspace{-0.6em}
\subsection{Experimental Details}
We follow their own code of the backbone models mentioned above but only modify the part of training sample construction\footnote{{\href{https://anonymous.4open.science/r/SETO-code-A026/}{The core source code: https://anonymous.4open.science/r/SETO-code-A026/}}}. 
To determine the value of the parameters for the sampling part of  RSS~\cite{RSS}, we conduct experiments using  \{0.3,0.5,0.7\} for $\tau$, which represents the maximum percentage of items in a sequence that can be selected as a target, and \{0.4,0.6,0.8\} for $\alpha$, which is analogous to the parameter in the probability function in Equation \ref{equ:probabilityFunction}.
To simulate diverse behavioral manifestations by our SETO, we set the $scope$ of the $Swap$ operation from the range \{0.2,0.4,0.6,0.8,1.0\}. The range of $\rho$ in the $Removal$ operation lies in \{0.1,0.3,0.5,0.7,0.9\}.
Other parameters such as the batch size, dropout rate, learning rate, hidden dimensions, and maximum length are respectively kept at default values of 128, 0.5, 0.001, 50, and 50 ~\cite{FISSA}. 
Due to space limitations, more experimental details can be found in Appendix~\ref{sec:details} in the auxiliary material.
% Note that our SETO\footnote{{\href{https://anonymous.4open.science/r/SETO-code-A026/}{The core source code: https://anonymous.4open.science/r/SETO-code-A026/}}} only temporarily handles the training sample part of each iteration and has nothing to do with the code of the model. 
 
\begin{table}[t]
\centering
   \caption {Recommendation performance of six backbone models (marked by *) and their variants with RSS and our SETO (including SETO(S) and SETO(R)). We bold the best results and underline the second-best ones. Note that although DuoRec, which improves on SASRec, is not a non-invasive approach, we also use it as a reference.}
  \vspace{-0.7em}
  \label{tab:SETO_baselines}
 \resizebox{1\linewidth}{!}{
\begin{tabular}{lcccccc}
\toprule
  & \multicolumn{2}{c}{Foursquare} & \multicolumn{2}{c}{Games} & \multicolumn{2}{c}{Beauty} \\
& \begin{tabular}[c]{@{}c@{}}Recall\\ @10\end{tabular} & \begin{tabular}[c]{@{}c@{}}NDCG\\ @10\end{tabular} & \begin{tabular}[c]{@{}c@{}}Recall\\ @10\end{tabular} & \begin{tabular}[c]{@{}c@{}}NDCG\\ @10\end{tabular} & \begin{tabular}[c]{@{}c@{}}Recall\\ @10\end{tabular} & \begin{tabular}[c]{@{}c@{}}NDCG\\ @10\end{tabular}\\ \midrule
Caser*  & 0.2518 & 0.1334 & 0.0546 & 0.0271 & 0.0188 & 0.0092 \\
  +RSS & {\ul 0.2733} & {\ul 0.1488} & 0.0728 & 0.0378 & {\ul 0.0228} & {\ul 0.0114} \\
 +SETO(S) & 0.2546 & 0.1360 & {\ul 0.0745} & {\ul 0.0383} & 0.0224 & 0.0113 \\
  +SETO(R) & \textbf{0.2797} & \textbf{0.1545} & \textbf{0.0853} & \textbf{0.0435} & \textbf{0.0228} & \textbf{0.0114} \\ \midrule
SR-GNN*  & 0.2593 & 0.1027 & 0.0766 & {\ul 0.0319} & 0.0403 & 0.0165 \\
  +RSS & 0.2635 & 0.1039 & 0.0707 & 0.0298 & 0.0357 & 0.0153 \\
  +SETO(S) & {\ul 0.2635} & {\ul 0.1041} & {\ul 0.0774} & 0.0316 & {\ul 0.0408} & \textbf{0.0167} \\
  +SETO(R) & \textbf{0.2642} & \textbf{0.1142} & \textbf{0.0812} & \textbf{0.0334} & \textbf{0.0410} & {\ul 0.0166} \\ \midrule
SASRec*   & 0.2413 & 0.1245 & 0.0767 & 0.0349 & 0.0373 & 0.0163 \\
DuoRec & {\ul 0.2650} & 0.1293 & 0.0810 & 0.0389 & 0.0346 & 0.0148 \\
  +RSS &  0.2462 & 0.1295 & 0.0797 & 0.0385 & 0.0359 & 0.0159 \\
  +SETO(S) & 0.2443 & {\ul 0.1297} & \textbf{0.0907} & \textbf{0.0403} & {\ul 0.0458} & {\ul 0.0197} \\
  +SETO(R) & \textbf{0.2710} & \textbf{0.1418} & {\ul 0.0856} & {\ul 0.0396} & \textbf{0.0462} & \textbf{0.0200} \\
 \midrule
FISSA*   & 0.2289 & {\ul 0.1215} & \textbf{0.0605} & {\ul 0.0289} & 0.0190 & 0.0088 \\
  +RSS & {\ul 0.2304} & 0.1209 & 0.0561 & 0.0287 & 0.0167 & 0.0081 \\
  +SETO(S) & 0.2063 & 0.1094 & 0.0518 & 0.0245 & {\ul 0.0232} & {\ul 0.0104} \\
  +SETO(R) & \textbf{0.2351} & \textbf{0.1230} & {\ul 0.0595} & \textbf{0.0296} & \textbf{0.0262} & \textbf{0.0135} \\ \midrule
FMLP-Rec*  & 0.2111 & 0.1188 & 0.0715 & 0.0372 & {\ul 0.0339} & {\ul 0.0174} \\
  +RSS & 0.2346 & 0.1354 & 0.0683 & 0.0349 & 0.0313 & 0.0166 \\
  +SETO(S) & {\ul 0.2392} & {\ul 0.1385} & \textbf{0.0776} & \textbf{0.0393} & \textbf{0.0355} & \textbf{0.0186} \\
  +SETO(R) & \textbf{0.2445} & \textbf{0.1388} & {\ul 0.0753} & {\ul 0.0383} & 0.0338 & 0.0172 \\ \midrule
  DiffuRec* & 0.2687 & 0.1538 & 0.1011 & 0.0527 & \textbf{0.0525} & \textbf{0.0291} \\
+RSS & 0.2716 & 0.1550 & 0.0896 & 0.0439 & 0.0429  & 0.0206  \\
+SETO(S) & {\ul 0.2756} & {\ul 0.1571} & \textbf{ 0.1049} & {\ul 0.0535} & {\ul 0.0522} & {\ul 0.0287} \\
+SETO(R) & \textbf{0.2786} & \textbf{0.1585} & {\ul 0.1042} & \textbf{0.0536} & 0.0519 & 0.0284\\
  \bottomrule
\end{tabular}
  \vspace{-0.5em}
}
\end{table}
\begin{table}[t]
\centering
\caption{Supplemental experiments of SASRec, FISSA and their variants with RSS and our SETO (including SETO(S) and SETO(R)) on the partial item set, which contains randomly sampled 100 items that the user has not interacted.}
  \vspace{-0.7em}
\label{tab:partial_candidate}
 \resizebox{1\linewidth}{!}{
\begin{tabular}{lcccccc}
\toprule
 & \multicolumn{2}{c}{Foursquare} & \multicolumn{2}{c}{Games} & \multicolumn{2}{c}{Beauty} \\
 & \begin{tabular}[c]{@{}c@{}}Recall\\ @10\end{tabular} & \begin{tabular}[c]{@{}c@{}}NDCG\\ @10\end{tabular} & \begin{tabular}[c]{@{}c@{}}Recall\\ @10\end{tabular} & \begin{tabular}[c]{@{}c@{}}NDCG\\ @10\end{tabular} & \begin{tabular}[c]{@{}c@{}}Recall\\ @10\end{tabular} & \begin{tabular}[c]{@{}c@{}}NDCG\\ @10\end{tabular} \\ \hline
SASRec*  & 0.4808 & 0.2611 & 0.6009 & 0.3685 & 0.3609 & 0.2214 \\
 +RSS & 0.4934 & 0.2727 & 0.5842 & 0.3586 & 0.3637 & 0.2144 \\
 +SETO(S) & \textbf{0.5151} & \textbf{0.2919} & {\ul 0.6104} & {\ul 0.3843} & \textbf{0.3872} & \textbf{0.2365} \\
 +SETO(R) & {\ul 0.5146} & {\ul 0.2865} & \textbf{0.6136} & \textbf{0.3747} & {\ul 0.3710} & {\ul 0.2337} \\ \midrule
FISSA*  & 0.5106 & 0.2794 & 0.6743 & 0.4134 & 0.4164 & 0.2484 \\
 +RSS & 0.4470 & 0.2353 & 0.6362 & 0.3699 & 0.4101 & 0.2436 \\
 +SETO(S) & {\ul 0.5277} & {\ul 0.2942} & {\ul 0.6824} & {\ul 0.4280} & \textbf{0.4361} & \textbf{0.2717} \\
 +SETO(R) & \textbf{0.5310} & \textbf{0.2942} & \textbf{0.6866} & \textbf{0.4325} & {\ul 0.4202} & {\ul 0.2584} \\ 
 \bottomrule
\end{tabular}
  \vspace{-0.7em}
}
\end{table}

% \vspace{-0.5em}
\section{Results}
\label{sec:results}
In this section, we show our experimental results to answer the questions presented in Section \ref{sec:rq}.

\begin{table}[]
\centering
\caption{Performance of MGCL~\cite{MGCL}, TJAPL~\cite{TJAPL} and their variants with our SETO (including SETO(S) and SETO(R)) on cross-domain sequential recommendation. Note that we show the results of the target domain while the other domains are the source domains (e.g., when the target domain is Movie, we use CD or Book as a source domain to assist in training) and only report the best performance.}
% \vspace{-0.7em}
 \resizebox{1\linewidth}{!}{
\begin{tabular}{lcccccc}
\toprule
 & \multicolumn{2}{c}{Movie} & \multicolumn{2}{c}{CD} & \multicolumn{2}{c}{Book} \\
 & \begin{tabular}[c]{@{}c@{}}Recall\\ @10\end{tabular} & \begin{tabular}[c]{@{}c@{}}NDCG\\ @10\end{tabular} & \begin{tabular}[c]{@{}c@{}}Recall\\ @10\end{tabular} & \begin{tabular}[c]{@{}c@{}}NDCG\\ @10\end{tabular} & \begin{tabular}[c]{@{}c@{}}Recall\\ @10\end{tabular} & \begin{tabular}[c]{@{}c@{}}NDCG\\ @10\end{tabular} \\ \midrule
MGCL* & 0.3784 & 0.2155 & 0.3903 & 0.2238 & 0.2804 & 0.1516 \\
+SETO(S) & {\ul 0.3896} & {\ul 0.2206} & {\ul 0.3961} & {\ul 0.2268} & \textbf{0.2950} & \textbf{0.1625} \\
+SETO(R) & \textbf{0.3918} & \textbf{0.2226} & \textbf{0.4037} & \textbf{0.2312} & {\ul 0.2929} & {\ul 0.1580} \\ \midrule
TJAPL* & 0.3698 & 0.2119 & 0.3810 & 0.2144 & 0.2961 & 0.1628 \\
+SETO(S) & \textbf{0.3872} & \textbf{0.2212} & \textbf{0.4028} & \textbf{0.2274} & \textbf{0.3095} & \textbf{0.1707} \\
+SETO(R) & {\ul 0.3752} & {\ul 0.2155} & {\ul 0.3901} & {\ul 0.2200} & {\ul 0.3027} & {\ul 0.1662}\\ \bottomrule
\end{tabular}
\label{tab:cross-domain}
  \vspace{-0.5em}
}
\end{table}
\begin{table}[]
\centering
 \caption{Relative improvement of a deployed recommendation method based on SASRec (denoted as Base) and its variants with our SETO (including SETO(S) and SETO(R)) on a large-scale industry dataset. 
 % Note that we show the improvement of each method over the Base model.
 }
  % \vspace{-0.7em}
  \label{tab:indus_data}
\begin{tabular}{lcc}
\toprule
Models     & Recall@50  & Recall@100   \\ \midrule
Base*       & 0\% & 0\%   \\
+ SETO(S)      & 1.19\% & 1.117\%  \\
+ SETO(R)      & 1.16\% & 0.73\%   \\  \bottomrule
\end{tabular}
  \vspace{-0.5em}
\end{table}

\subsection{RQ1: Performance Comparison}
\label{subsec:Performance Comparison}
We employ the closely related method RSS~\cite{RSS} and two variants (i.e., $Swap$ and $Removal$) of our SETO to six backbone models, and the results are presented in Table \ref{tab:SETO_baselines}.  
In our SETO, we transform both the input subsequence and target subsequence via $Swap$ and $Removal$, respectively, denoted by `SETO(S)' and `SETO(R)'.

For a comparison of the recommendation performance of the models themselves, we find that DiffuRec performs best, which is the same as in the report from~\cite{DiffuRec}.
The session-based model SR-GNN performs better on datasets with shorter sequences, i.e., Foursquare and Beauty. And on Games, SASRec performs better.
We find that FISSA does not perform well here, and speculate the reason is that it is one of the few models that adds candidate items to model training, which does not show its strengths on the full candidate set. 
In order to show the effectiveness of our SETO more comprehensively, we follow FISSA~\cite{FISSA} and supplement the experiments on the partial candidate set in Table~\ref{tab:partial_candidate}.

Overall, our SETO significantly enhances the recommendation performance of all six backbone models on most datasets except DiffuRec on Beauty.
In fact, DiffuRec is a method of adding randomness and uncertainty to training~\cite{DiffuRec}, injecting further randomness can affect the final result, especially on the very sparse dataset, i.e., Beauty. However, there is a boost in both Foursquare and Games.
Compared with the sequence-to-item models, sequence-to-sequence models (i.e., SASRec and FISSA on the partial candidate set) can be enhanced by our SETO more. For FMLP-Rec with denoising, it counteracts our enhancement effect to a certain extent, so the boost is not obvious.
% Whereas the augmentation method RSS enhances the model's recommendation performance on denser datasets, i.e., Foursquare, it is unstable on sparse datasets.
Consistent with \cite{RSS}, using RSS alone without the designed loss function $\lambda Rank$ does not yield stably effective boosting results.
As for the model-dependent sequence augmentation method DuoRec, it enhances SASRec on Foursquare and Games, but not as much as our SETO.

In order to adapt to different datasets and different models, we design two different operations, i.e., $Swap$ and $Removal$. 
$Removal$ is more suitable for datasets that are more sensitive to the backward and forward order of the items in a sequence, e.g., Foursquare, because it does not change the order of the items in the sample sequences. 
For a sparse dataset, the advantage of $Swap$'s ability to enrich a wider variety of samples can be demonstrated,  e.g., the performance of FMLP-Rec on Games and Beauty.

% \vspace{-0.5em}
\subsection{RQ2: Performance on Cross-Domain Sequential Recommendation}
We apply our SETO to the cross-domain sequential recommendation scenario to prove the highly generic feature. 
We take MGCL~\cite{MGCL} and TJPAL~\cite{TJAPL} as the backbone models, which utilize GNNs and some attention mechanisms to learn the information of different domains, respectively. 
The results of MGCL and TJAPL in Table~\ref{tab:cross-domain} are close with those reported in ~\cite{MGCL,TJAPL}.
Moreover, the results verify that our SETO can be easily applied to a model focusing on a single target domain (i.e., MGCL) and that both source and target domains (i.e., TJPAL), and help them to get better performance. 
Since there are more input sequences in TJAPL, applying $Removal$ to every sequence will result in too much information loss. 
Hence, $Removal$ is slightly less effective than $Swap$ but is still helpful.

\begin{figure*}[]
\centering
   \begin{minipage}[t]{0.32\linewidth}
   \centerline{(a) Foursquare}
     \begin{minipage}[t]{0.48\linewidth}
        \centering
        \includegraphics[width=\textwidth,trim =0  6cm  0 0]{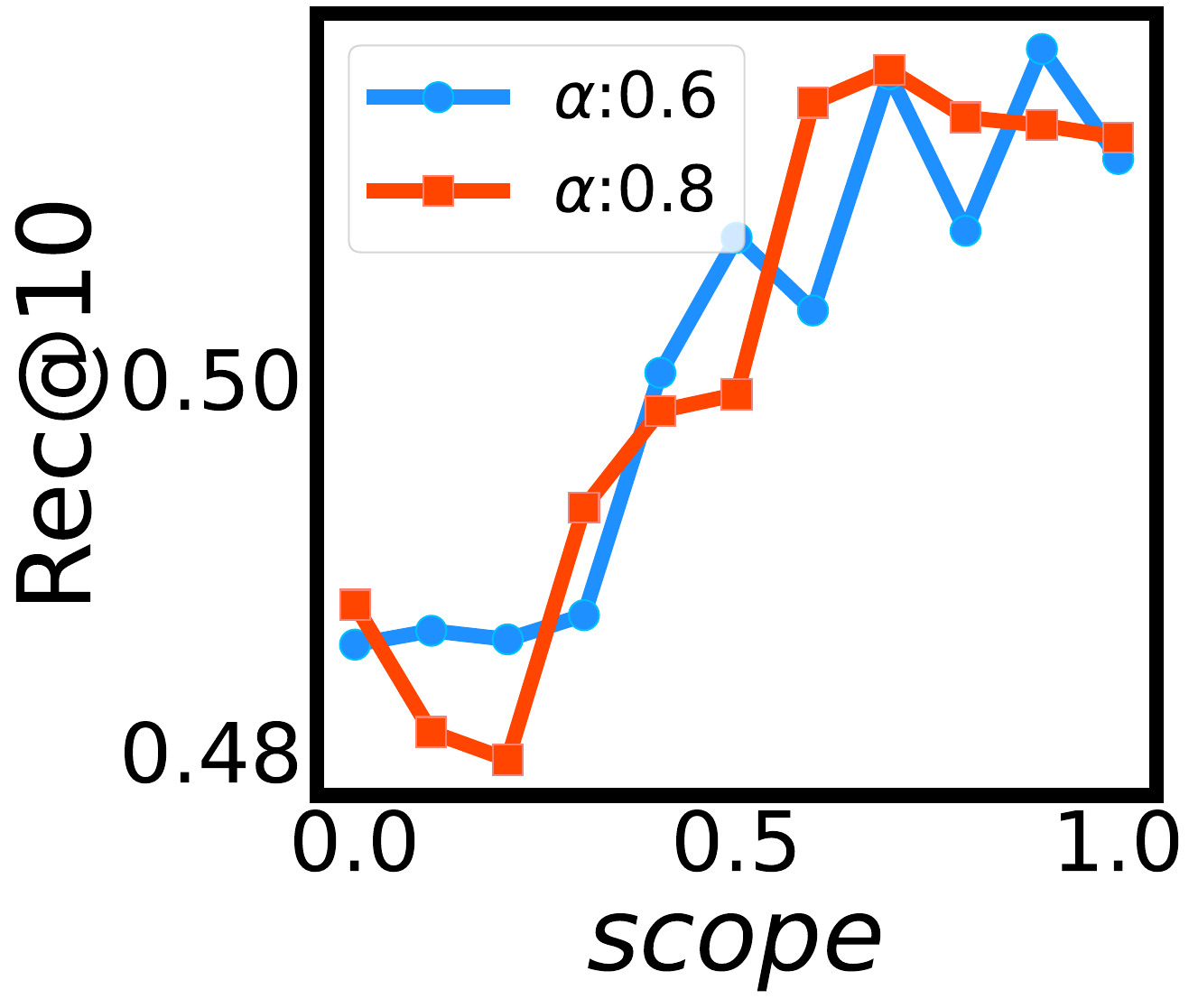}
        \label{subfig:Foursquare_scop}
    \end{minipage}%
    \begin{minipage}[t]{0.48\linewidth}
        \centering
        \includegraphics[width=\textwidth,trim =0  6cm  0 0]{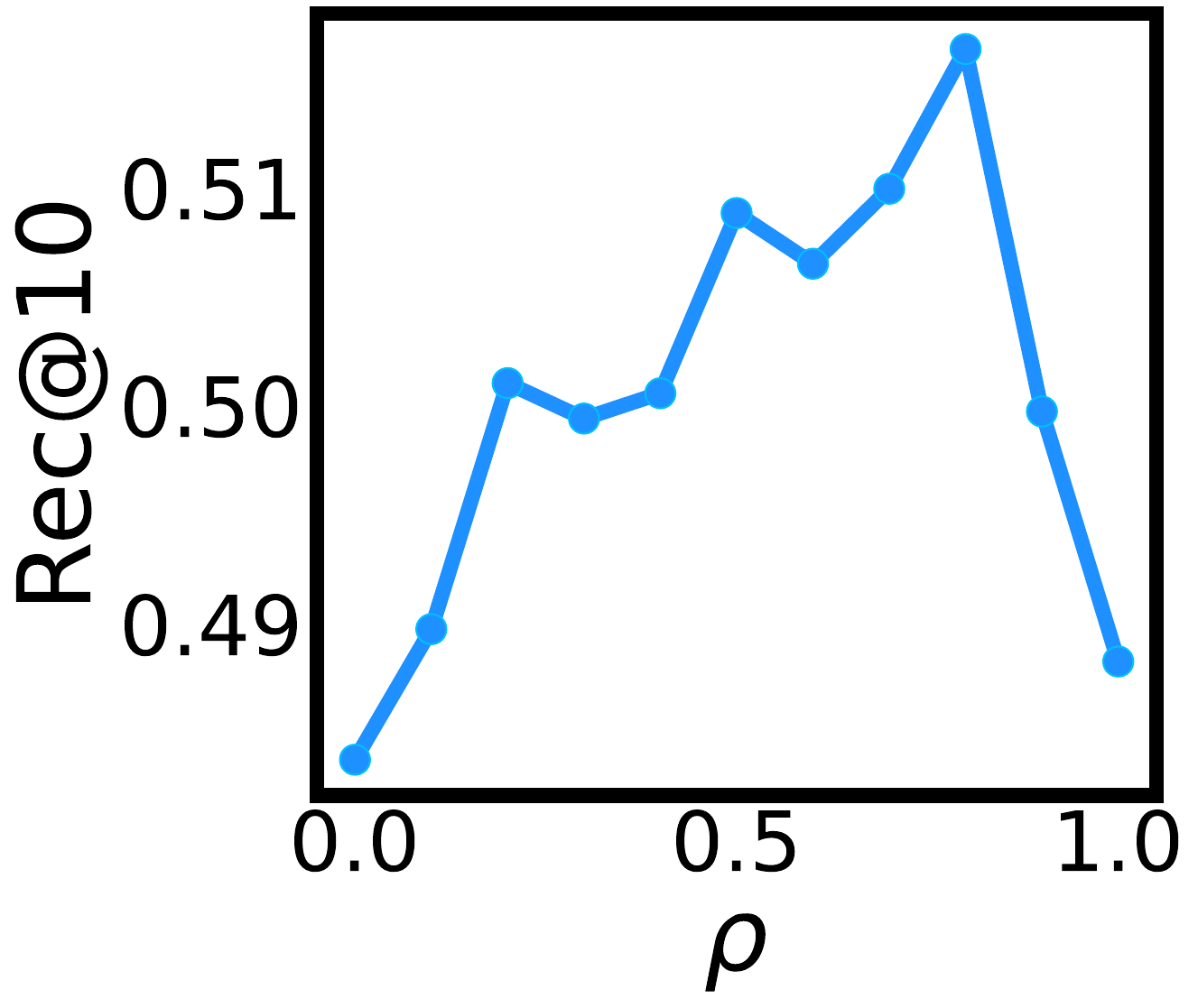}
        \label{subfig:Foursquare_rho}
    \end{minipage}
    \end{minipage}
\begin{minipage}[t]{0.32\linewidth}
    \centerline{(b) Games}
     \begin{minipage}[t]{0.48\linewidth}
        \centering
        \includegraphics[width=\textwidth,trim =0  6cm  0 0]{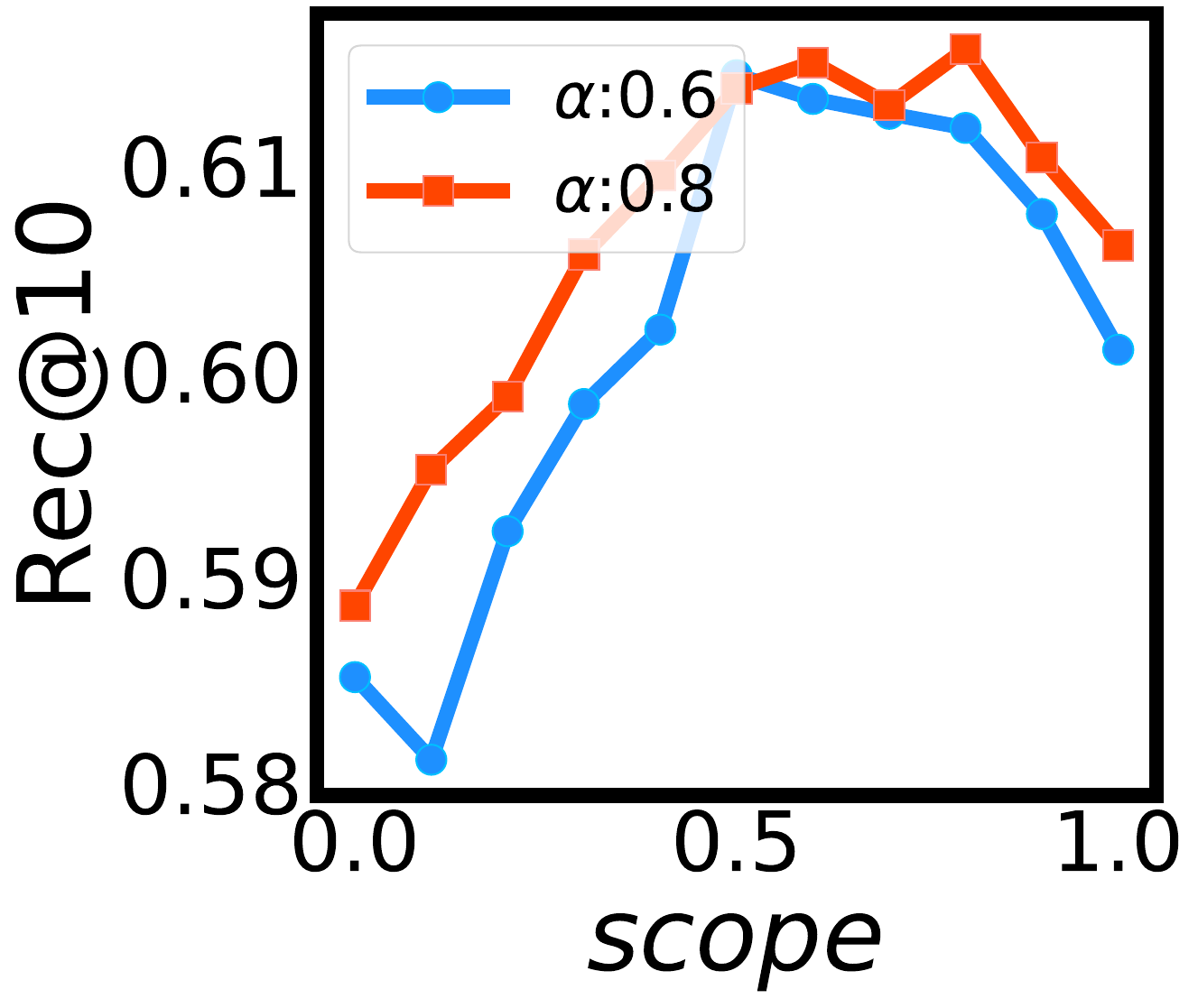}
        \label{subfig:Games_scope}
    \end{minipage}%
    \begin{minipage}[t]{0.48\linewidth}
        \centering
        \includegraphics[width=\textwidth,trim =0  6cm  0 0]{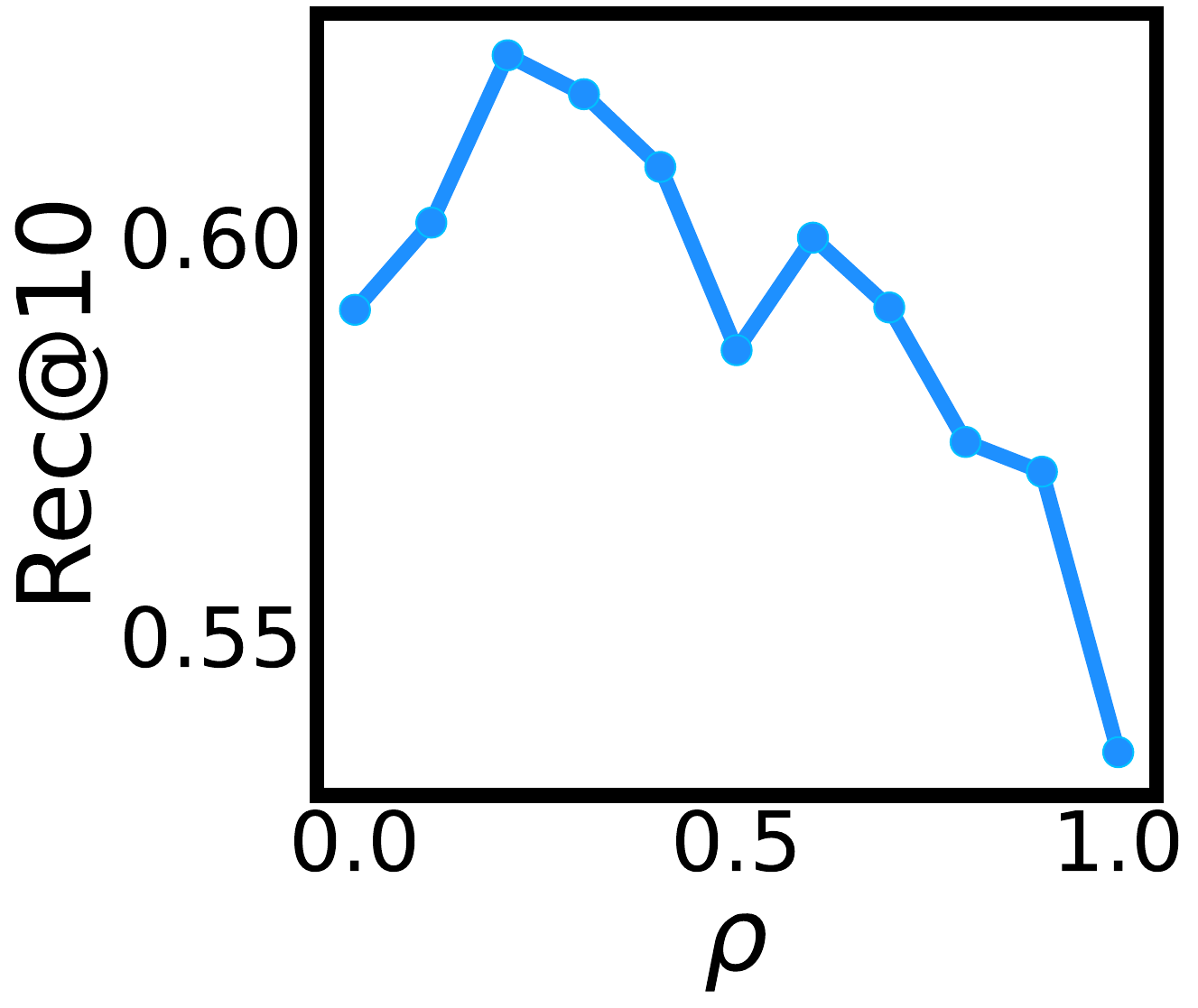}
        \label{subfig:Games_rho}
    \end{minipage}
\end{minipage}
\begin{minipage}[t]{0.32\linewidth}
        \centerline{(c) Beauty}
     \begin{minipage}[t]{0.48\linewidth}
        \centering
        \includegraphics[width=\textwidth,trim =0  6cm  0 0]{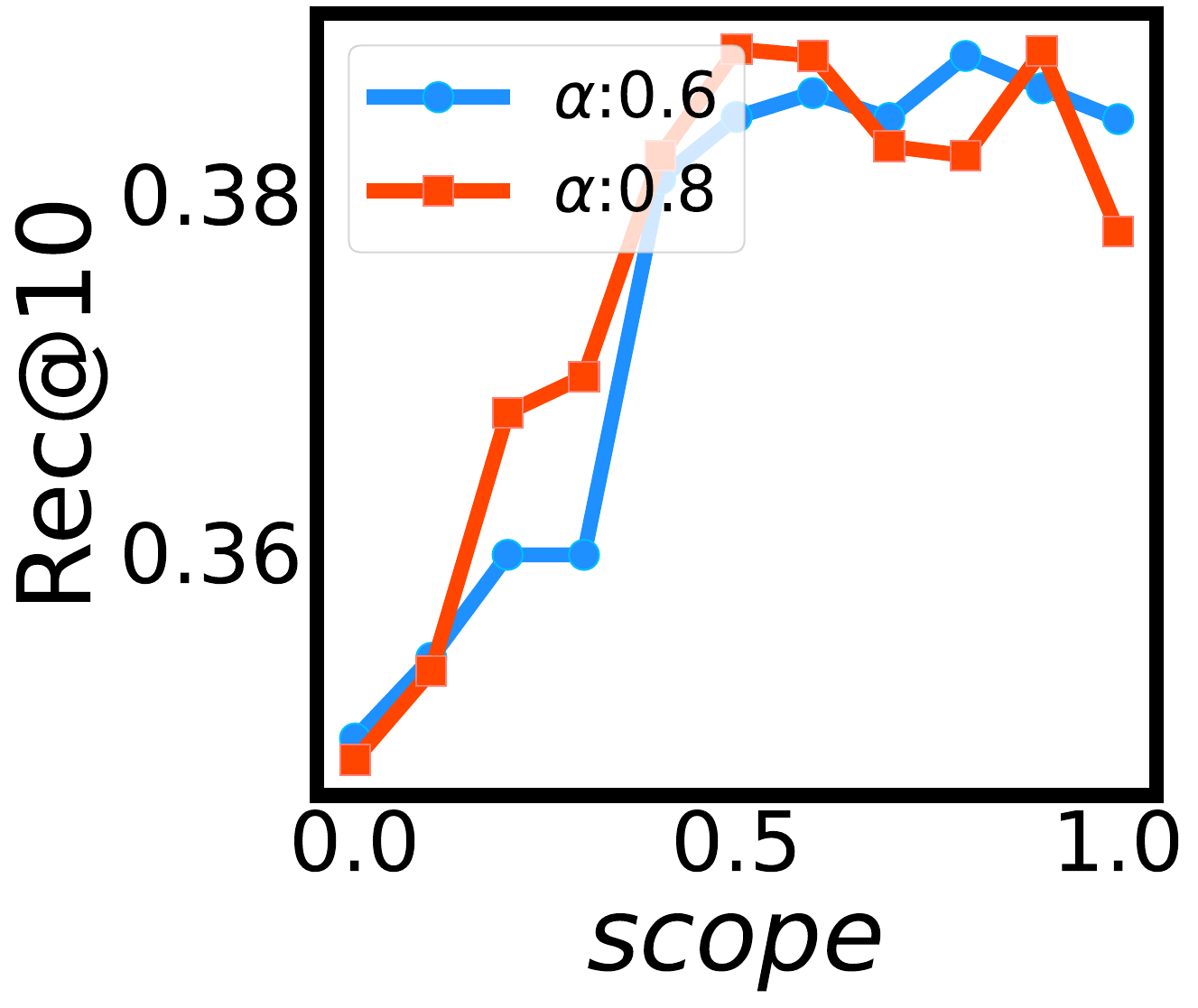}
        \label{subfig:Bearty_scope}
    \end{minipage}%
    \begin{minipage}[t]{0.48\linewidth}
        \centering
        \includegraphics[width=\textwidth,trim =0  6cm  0 0]{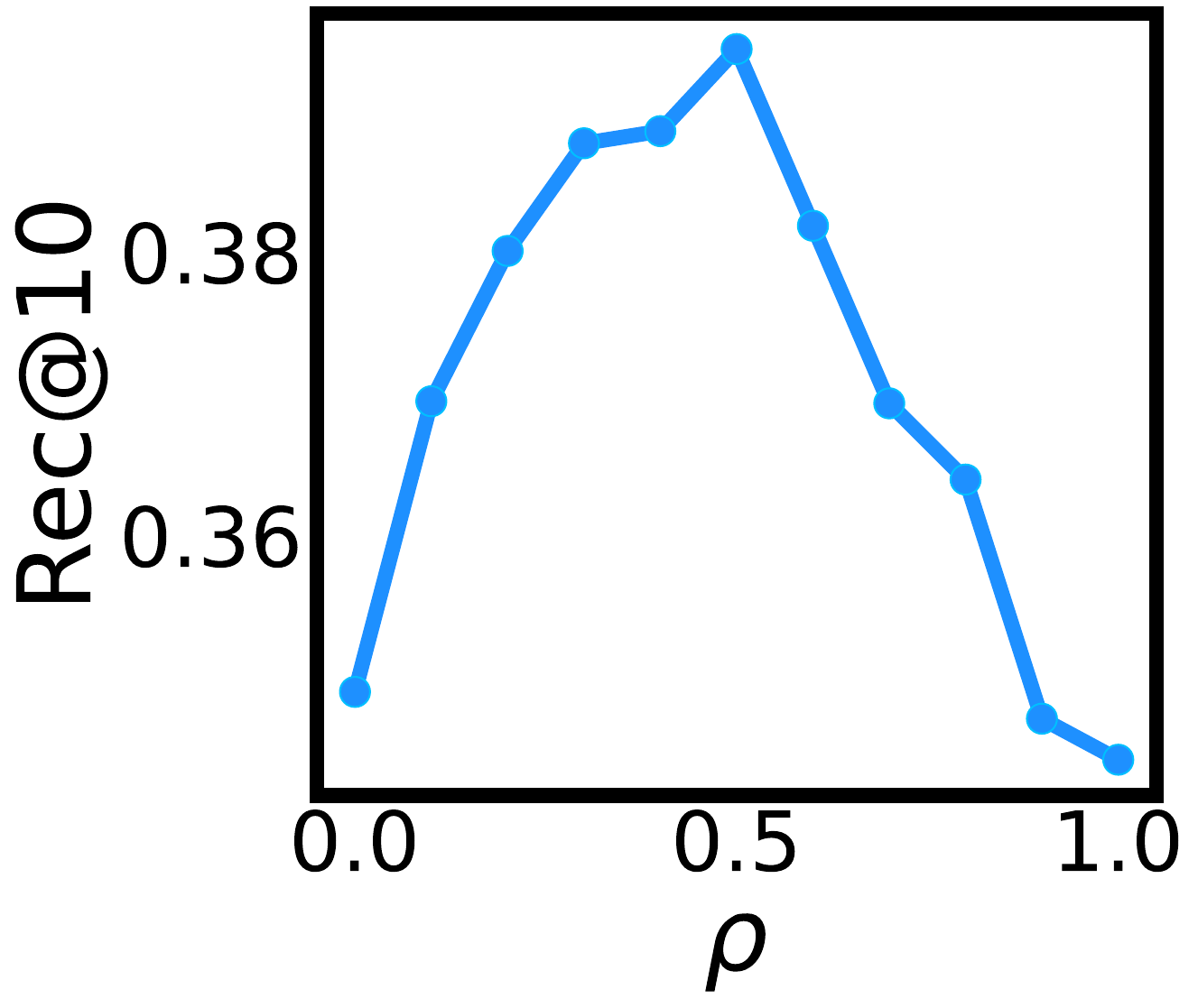}
        \label{subfig:Bearty_rho}
    \end{minipage}
\end{minipage}
\centering
   \begin{minipage}[t]{0.32\linewidth}
    \vspace{1em}
     \begin{minipage}[t]{0.48\linewidth}
        \centering
        \includegraphics[width=\textwidth,trim =0  6cm  0 0]{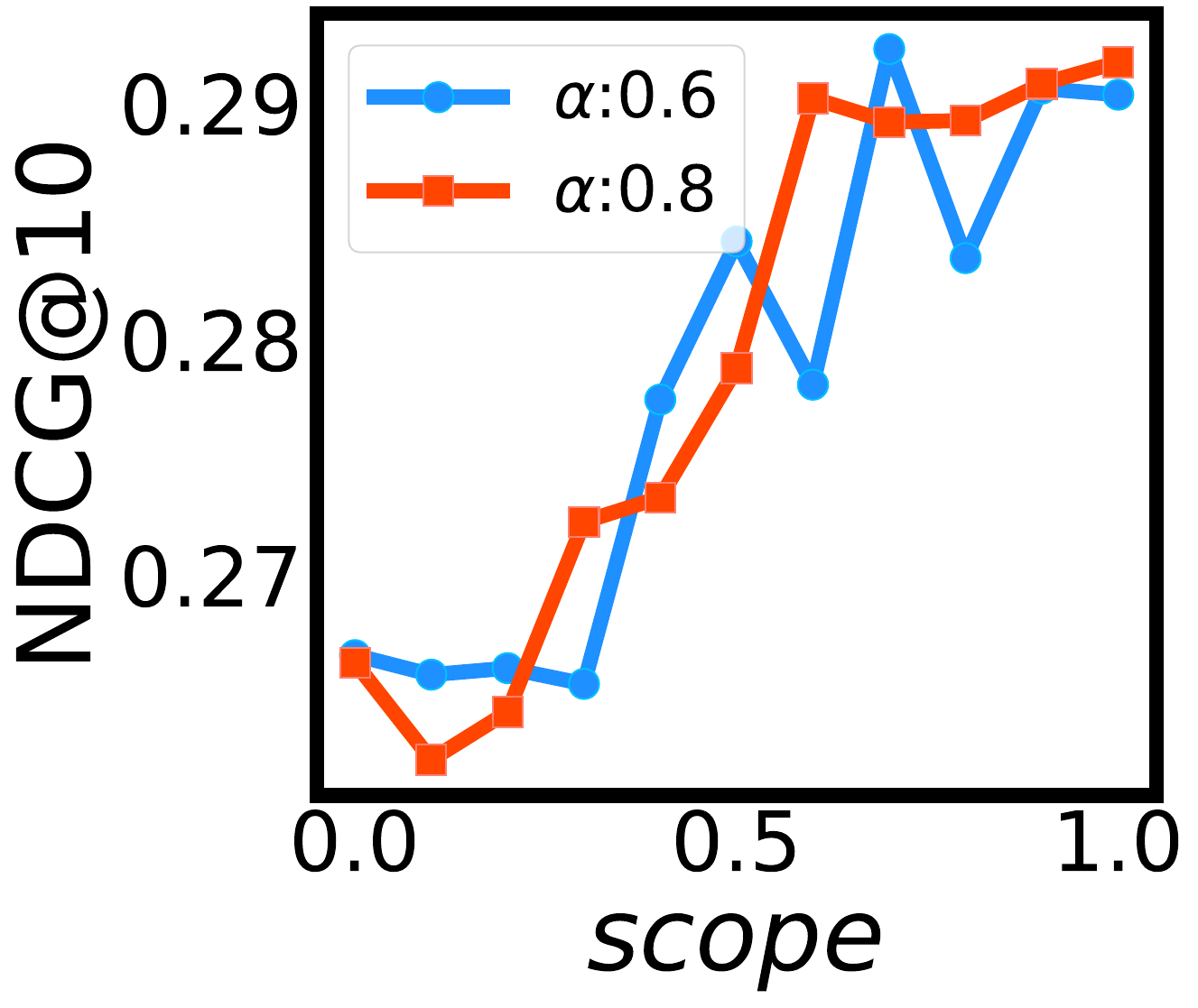}
        \label{subfig:Foursquare_scope_ndcg}
    \end{minipage}%
    \begin{minipage}[t]{0.48\linewidth}
        \centering
        \includegraphics[width=\textwidth,trim =0  6cm  0 0]{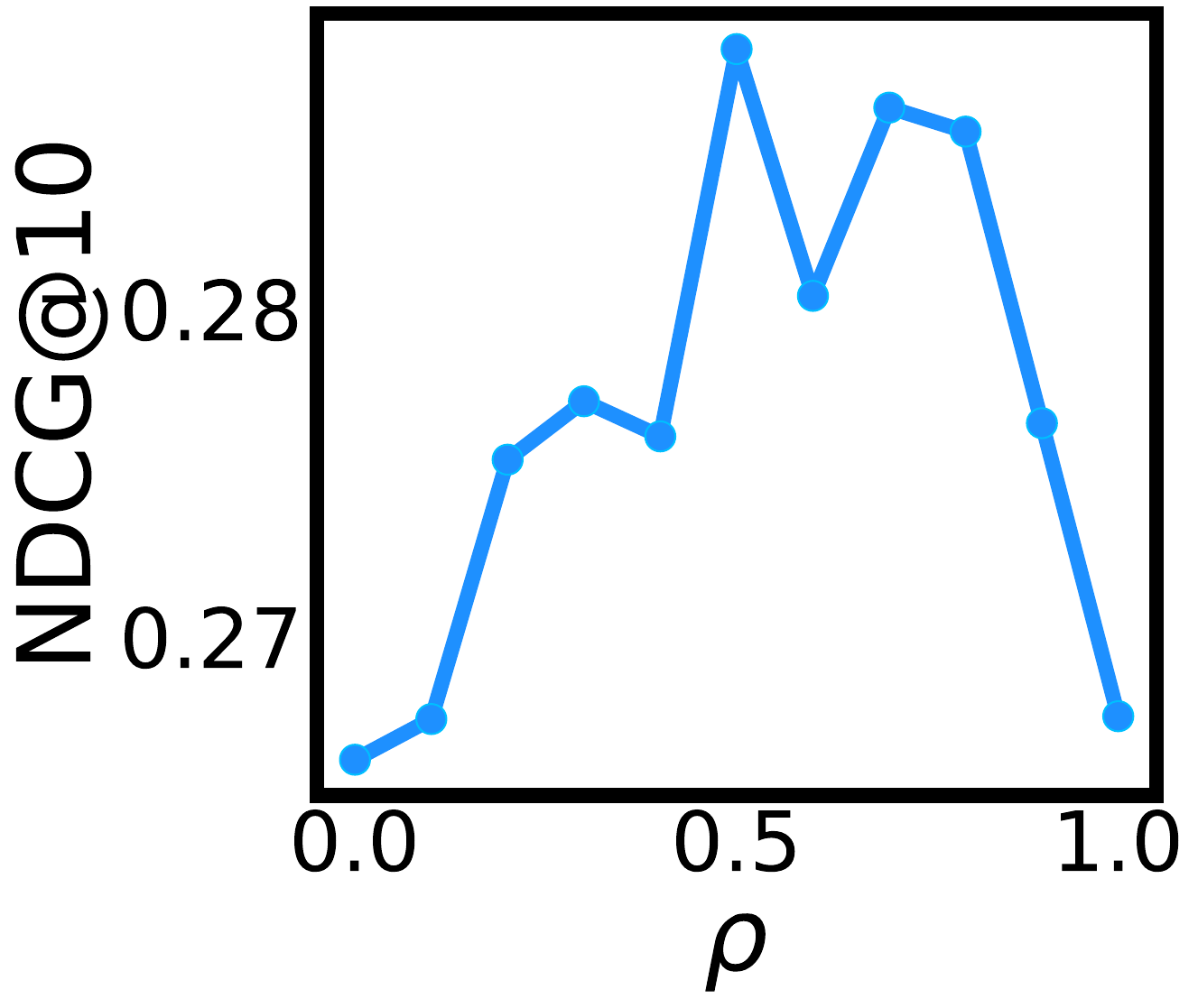}
        \label{subfig:Foursquare_rho_ndcg}
    \end{minipage}
    \end{minipage}
\begin{minipage}[t]{0.32\linewidth}
     \vspace{1em}
     \begin{minipage}[t]{0.48\linewidth}
        \centering
        \includegraphics[width=\textwidth,trim =0  6cm  0 0]{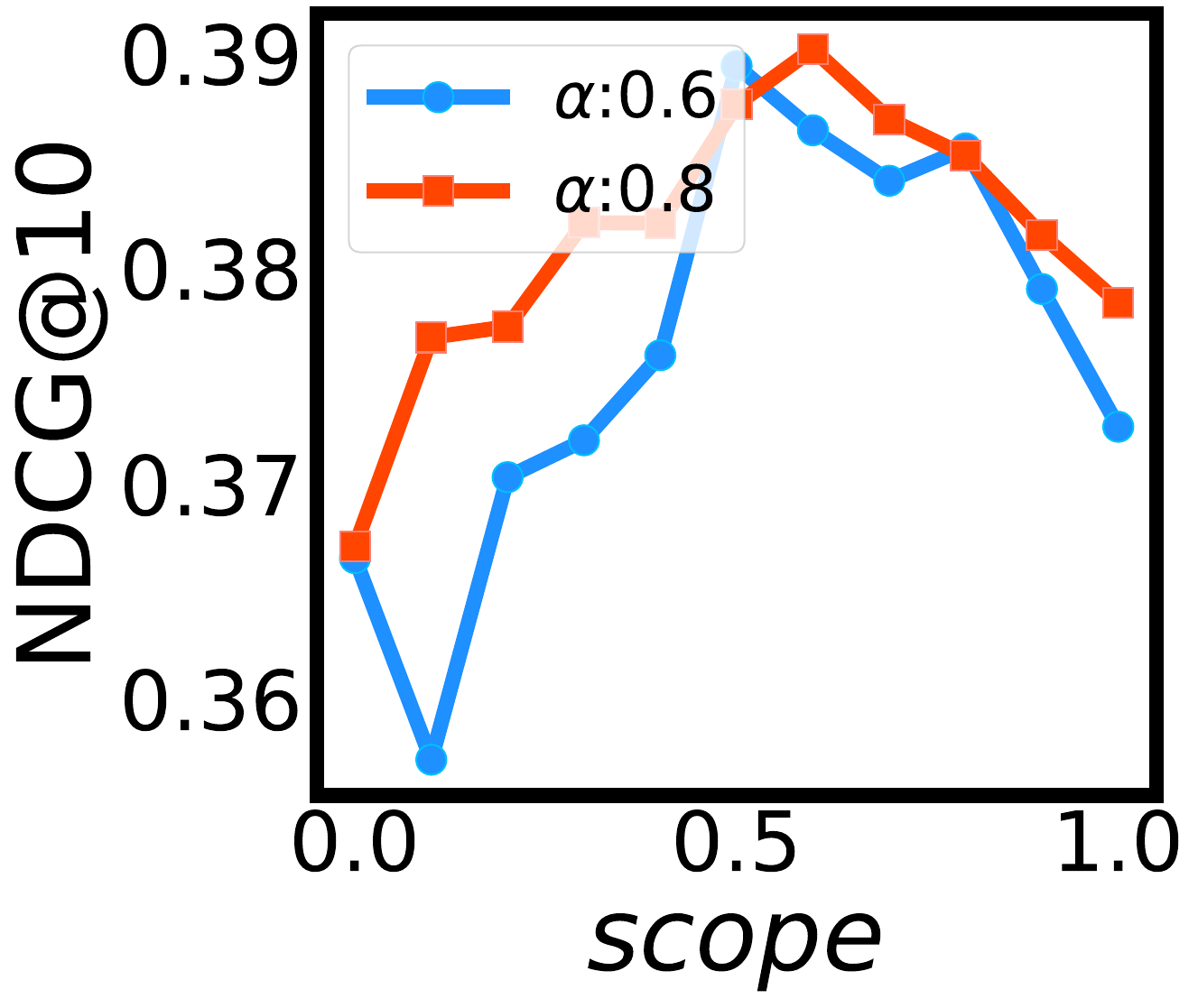}
        \label{subfig:Games_scope_ndcg}
    \end{minipage}%
    \begin{minipage}[t]{0.48\linewidth}
        \centering
        \includegraphics[width=\textwidth,trim =0  6cm  0 0]{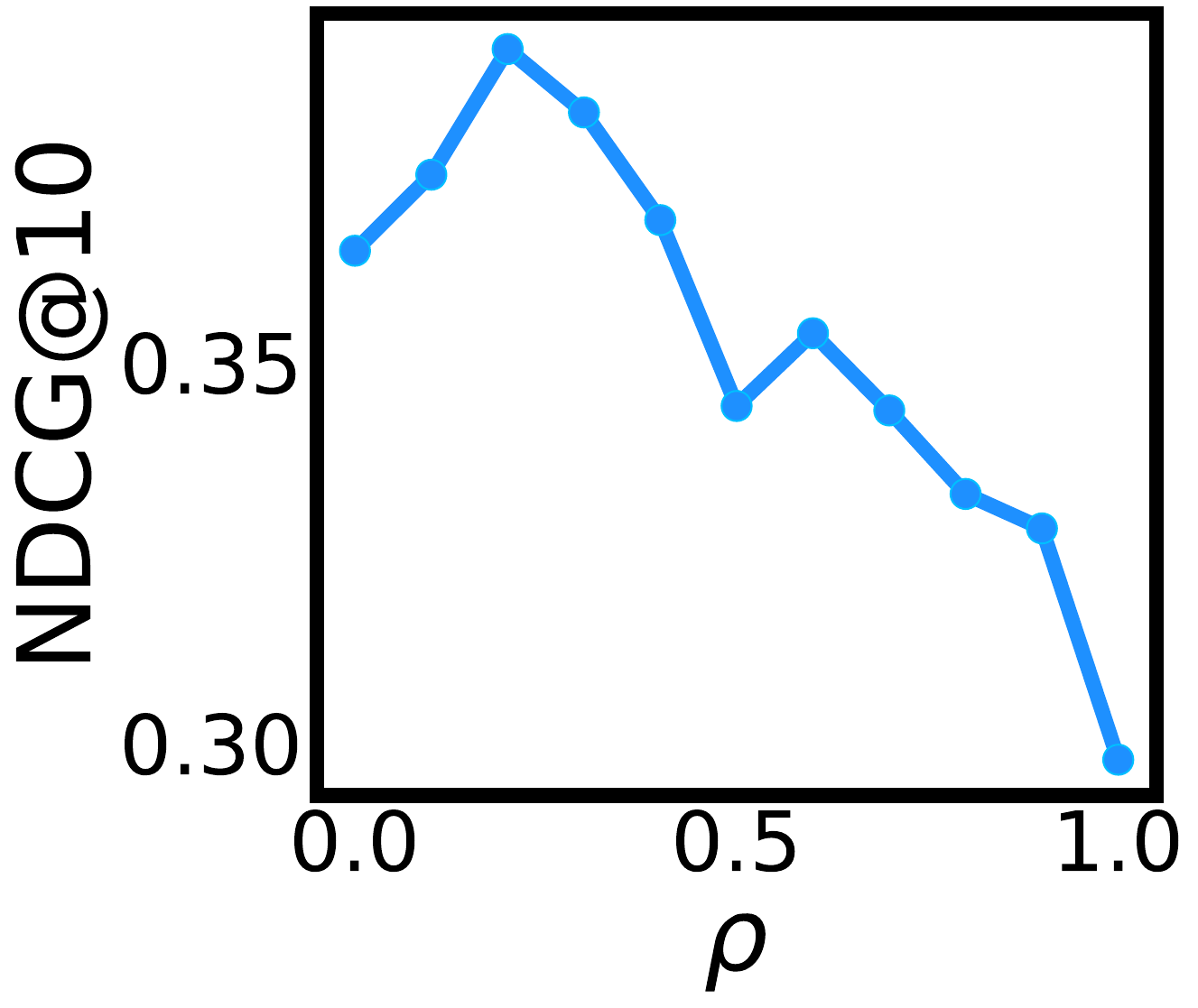}
        \label{subfig:Games_rho_ndcg}
    \end{minipage}
\end{minipage}
\begin{minipage}[t]{0.32\linewidth}
         \vspace{1em}
     \begin{minipage}[t]{0.48\linewidth}
        \centering
        \includegraphics[width=\textwidth,trim =0  6cm  0 0]{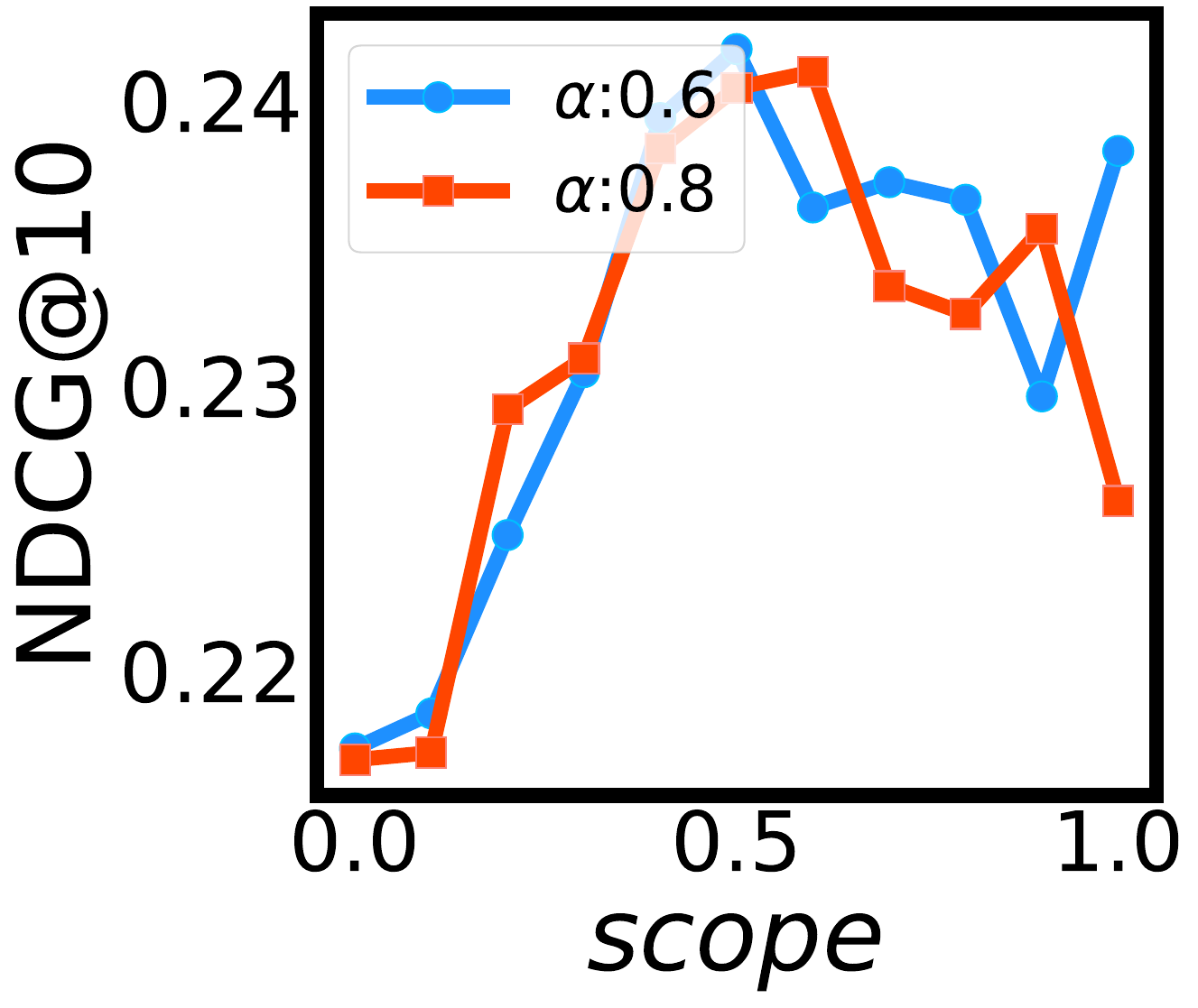}
        \label{subfig:Bearty_scope_ndcg}
    \end{minipage}%
    \begin{minipage}[t]{0.48\linewidth}
        \centering
        \includegraphics[width=\textwidth,trim =0  6cm  0 0]{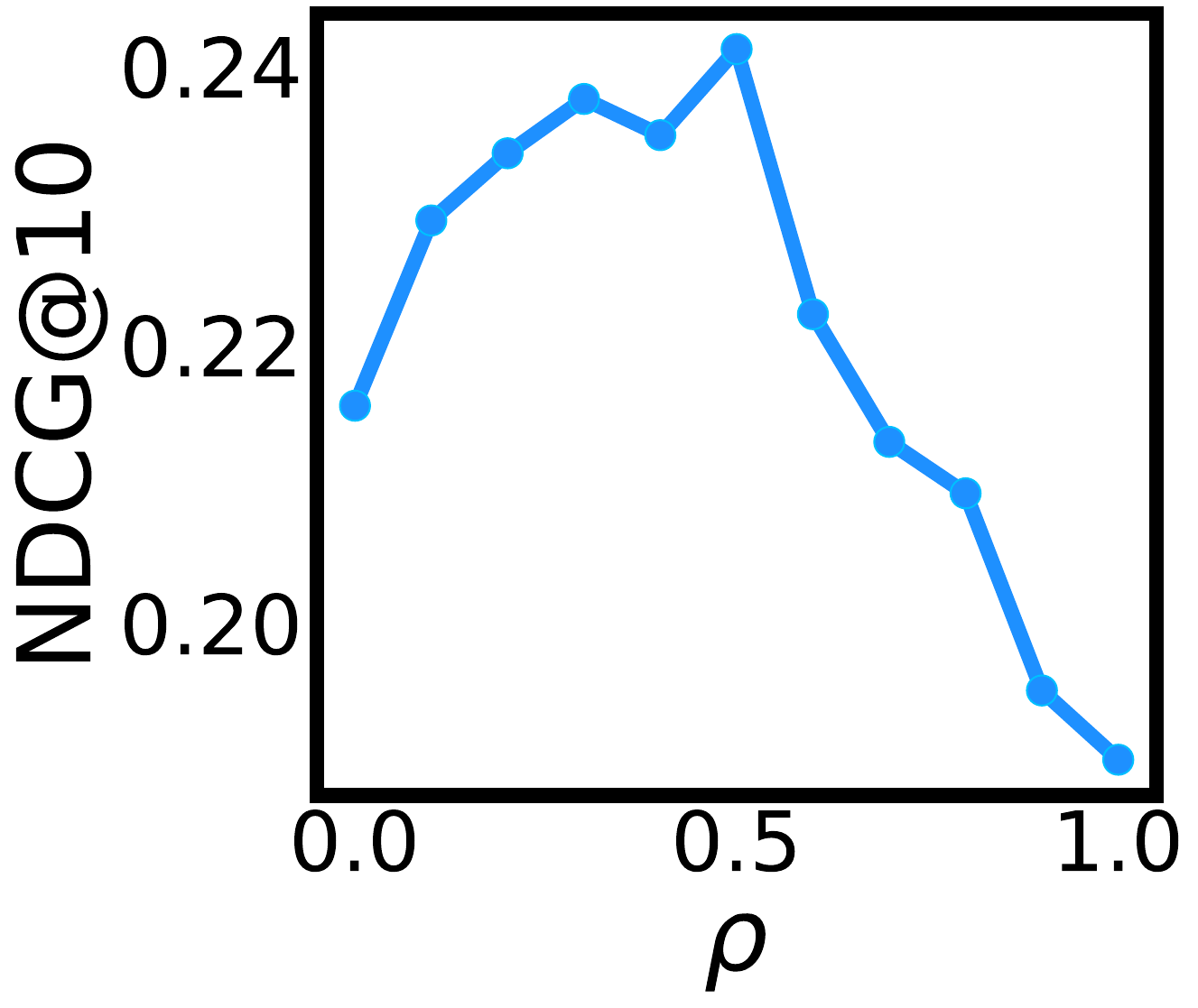}
        \label{subfig:Bearty_rho_ndcg}
    \end{minipage}
\end{minipage}
% \vspace{1em}
    \caption{Recommendation performance of SASRec via our SETO(S) using different values of $scope$ and $\rho$ on three datasets. Note that the y-axis starting point does not start from 0.}
    \label{fig:Hyper-parameters}
\end{figure*}

% \vspace{-0.5em}
\subsection{RQ3: Performance on Industry Data}
\label{sec:Q4}
To verify that our proposed SETO enriches more training samples and overcomes the challenge of data sparsity as a non-intrusive sequence augmentation method, we investigate its performance on a large-scale industry dataset, which consists of users' listening logs from a popular music streaming platform. 
Specifically, we use 10 days' worth of samples for training and the samples from the following day for test, resulting in training and test sets containing approximately 80 million and 8 million samples, respectively. 
RSS is not very friendly to be applied to a complex scenario as it requires that the input subsequence and the target subsequence to be non-repeating in two parts.
The model Base is a deployed sequential recommendation method based on SASRec~\cite{SASRec}. We evaluate the effectiveness of our SETO as a non-intrusive sequence augmentation method using Recall@50 and Recall@100 following industry practice, where the experimental results are presented in Table \ref{tab:indus_data}.
The results show that both the $Removal$ and $Swap$ variants of our SETO can improve the performance on Recall@50 by 1.1\%, relative to the baseline model. 
The improvement is evident and very promising given the large scale of the dataset and the fact that we did not make any changes to the model architecture or loss function. 
% The results again show that our approach is able to leverage the randomness inherent in users' sequences to provide more training opportunities and overcome the challenge of data sparsity.

% \vspace{-0.5em}
\subsection{RQ4: Ablation Study}
\label{subsec:RQ2}
In this section, we conduct ablation experiments on three different cases (i.e., $S_i, S_t, S_{i+t}$) on SASRec to answer RQ4 and present the results in Table \ref{tab:AblationStudy}.
$S_i$ (or $S_t$) denotes only processing the input(or taget) subsequences, respectively. $S_{i+t}$ denotes processing both the input and target subsequences.
Additionally, we compare the performance between our SETO and the completely random operations that are denoted by `Random(S)' and `Random(R)', respectively. 
`Random(S)' indicates that two items are randomly selected to swap without any limitation of scope and probability.
Similarly, `Random(R)' denotes that the number of removals and the items to be removed are also randomly selected without any limitation of scope. 
In a word, `Random(S)' and `Random(R)' exclude all plausibility constraints in sequence enhancement operations and retain only randomness.
\begin{table}[t]
\centering
\caption{ Performance comparison between SASRec and our SETO on different subsequences, i.e., only processing the input subsequences, only processing the target subsequences and processing both the input and target subsequences.}
  % \vspace{-0.7em}
 \label{tab:AblationStudy}
 \resizebox{1\linewidth}{!}{
\begin{tabular}{llcccccc}
\toprule
\multicolumn{2}{l}{} & \multicolumn{2}{c}{Foursquare} & \multicolumn{2}{c}{Games} & \multicolumn{2}{c}{Beauty} \\
\multicolumn{2}{l}{} & \begin{tabular}[c]{@{}c@{}}Recall\\ @10\end{tabular} & \begin{tabular}[c]{@{}c@{}}NDCG\\ @10\end{tabular} & \begin{tabular}[c]{@{}c@{}}Recall\\ @10\end{tabular} & \begin{tabular}[c]{@{}c@{}}NDCG\\ @10\end{tabular} & \begin{tabular}[c]{@{}c@{}}Recall\\ @10\end{tabular} & \begin{tabular}[c]{@{}c@{}}NDCG\\ @10\end{tabular} \\ \midrule
\multicolumn{2}{l}{SASRec} & 0.2413 & 0.1245 & 0.0767 & 0.0349 & 0.0373 & 0.0163 \\ \midrule
+Random(S) & $S_{i+t}$ & 0.2355 & 0.1271 & 0.0869 & 0.0392 & 0.0443 & 0.0190 \\ \midrule 
\multirow{3}{*}{+SETO(S)} & $S_{i}$ & 0.2403 & 0.1221 & 0.0867 & 0.0386 & 0.0450 & 0.0195 \\
 & $S_{t}$ & 0.2378 & 0.1281 & {\ul 0.0884} & 0.0393 & 0.0447 & 0.0192 \\
 & $S_{i+t}$ & 0.2443 & 0.1297 & \textbf{0.0907} & \textbf{0.0403} & {\ul 0.0458} & {\ul 0.0197} \\ \midrule
+Random(R) & $S_{i+t}$ & 0.2261 & 0.1177 & 0.0800 & {\ul 0.0402} & 0.0348 & 0.0169 \\ \midrule
\multirow{3}{*}{+SETO(R)} & $S_{i}$ & \textbf{0.2710} & \textbf{0.1418} & 0.0825 & 0.0369 & 0.0334 & 0.0142 \\
 & $S_{t}$ & 0.2414 & 0.1292 & 0.0842 & 0.0373 & \textbf{0.0462} & \textbf{0.0200} \\
 & $S_{i+t}$ & {\ul 0.2484} & {\ul 0.1324} & 0.0856 & 0.0396 & 0.0405 & 0.0183\\ \bottomrule
\end{tabular}
}
% \vspace{-1.5em}
\end{table}
% Note that the best and second-best results are marked in bold and italic, respectively. 
Complete randomization does not achieve good results, which proves the rationality and effectiveness of our design that combines probabilistic and randomized choices.
Clearly, applying our SETO to either a single subsequence or both subsequences already achieves significant improvement. 
For $Swap$, we find that processing both the input and target subsequences can achieve better results than separately one subsequence. 
This is because it can enrich more types of training samples.
However, for $Removal$, it is not as good as separately processing one subsequence. 
This can be attributed to the advantages mentioned in Section~\ref{subsec:Discussion}.
Only removing the items on a separate side does not cause the removed items to disappear in their entirety in a pair of training samples, but applying $Removal$ on both the input and target subsequences may cause loss of certain items' information, although it contains randomization.

% \vspace{-0.5em}
\subsection{RQ5: Effectiveness of Hyperparameters}
In this section, we examine the impact of various values of $scope$  on the prediction performance of the $Swap$ operation, as well as the influence of different values of $\rho$ in the $Removal$ operation on SASRec, with the other parameters fixed. 
% In fact, the trends exhibited for different values of the hyperparameters are very similar on the two evaluation metrics, i.e., Recall@10 and NDCG@10, and are more evident for the partial candidate sets. 
% Therefore, due to space constraints, we show the trend of NDCG@10 on the partial candidate sets in Figure \ref{fig:Hyper-parameters}.
The parameter $\alpha$ in Equation \ref{equ:probabilityFunction} affects the value of the probability function and thus affects the selection of items to be swapped. However, this can be considered as an extension to the $Swap$ method, so here we only try two values of $\alpha$ to observe the final experimental results of $scope$.
The results are shown in Figure~\ref{fig:Hyper-parameters}.

These two hyperparameters are designed to be more adaptable to subsequences of different lengths in the form of percentages than fixed values.
For $scope$, the datasets with more short sequences, i.e., Foursquare, larger value achieves better results, while Games and Beauty do well at 0.5. This suggests that it is reasonable to set a hyperparameter specifying the farthest position of $Swap$.
Additionally, the hyperparameter $\rho$ can determine the maximum number of items to be removed from a subsequence.
A similar trend is shown on all three datasets, i.e., rise first, then fall.
As a result, too large a value of $\rho$ removes too much information about the item, while too small a value is not conducive to enriching the sample. 
Overall, by selecting appropriate parameter values based on the sequence information of each dataset, we can bring these temporarily enriched samples closer to the fact in our SETO. 
% \vspace{-0.5em}
\section{Conclusions and Future Work}
In this paper, we propose a novel model-agnostic and generic framework called sample enrichment via temporary operations on subsequences (SETO), which fills the transformation space between the observed data and the underlying preferences with randomness and enriches training samples by applying two operations with rationality constraints to subsequences from both the input and output perspectives.
 Through extensive experiments on six real-world datasets, we show that our SETO can help various representative and state-of-the-art single-domain and cross-domain sequential recommendation methods and improve their performance significantly.
Moreover, empirical studies on a large industry dataset also show the effectiveness of our SETO in delivering significantly more accurate recommendations. 
Because of the non-intrusive nature of our SETO, we are interested in applying it to more scenarios.

\bibliographystyle{ACM-Reference-Format}
\bibliography{sample-base}

% \newpage
\newpage
\newpage
\appendix
% \balance
% \textcolor{blue}{
In this appendix, we first provide a comparison with traditional enhancement strategies to demonstrate the novelty of our SETO in Appendix~\ref{sec:vs}.
We also provide pseudocode for our SETO in Appendix~\ref{sec:algorithm}, including the overall algorithm in the sampler and the algorithm for sequence augmentation operations (i.e., $Swap$ and $Removal$) with rationality constraints.
Then, we introduce the backbone models adopted in the experiments in Appendix~\ref{sec:backbone} and supplement the setup details of the experimental hyperparameters in Appendix~\ref{sec:details}.
After that, we analyze the training efficiency of our SETO and other related methods in Appendix~\ref{sec:efficiency}.
Finally, we include discussion about the Amazon dataset in Appendix~\ref{sec:Amazon}.
% }

\begin{figure*}[t]
    \centering
    \includegraphics[width=0.33\linewidth]{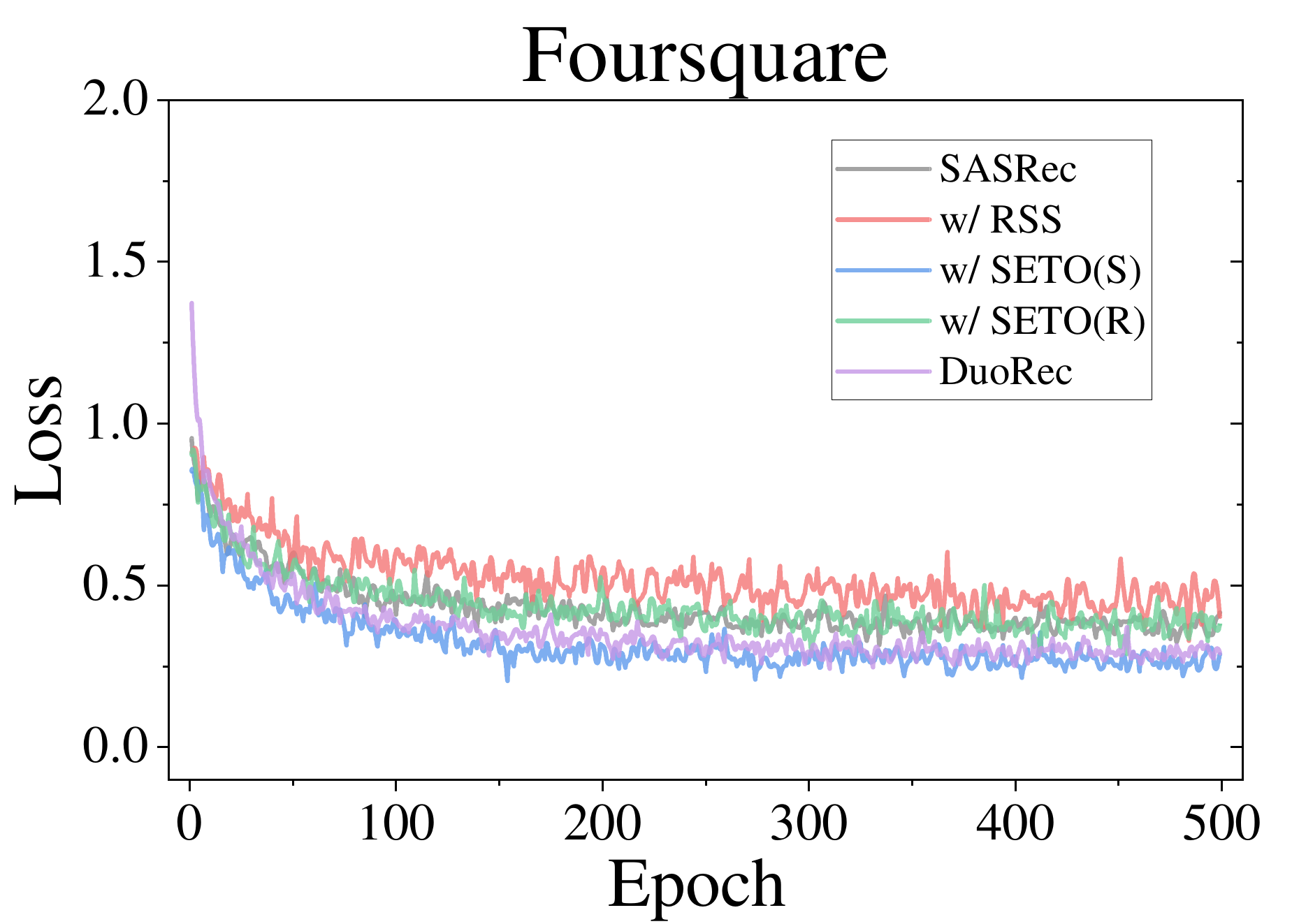}
    \includegraphics[width=0.33\linewidth]{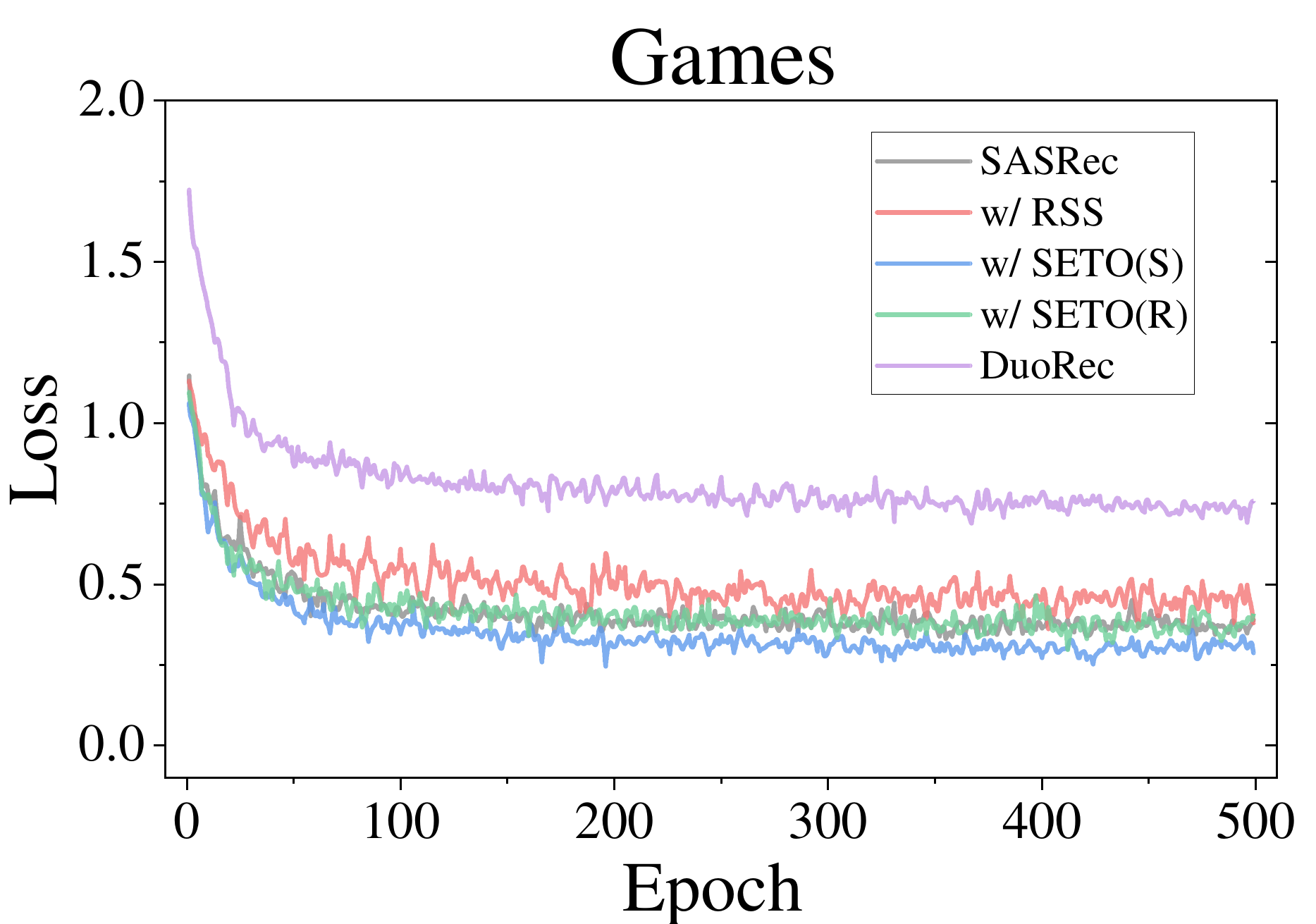}
    \includegraphics[width=0.33\linewidth]{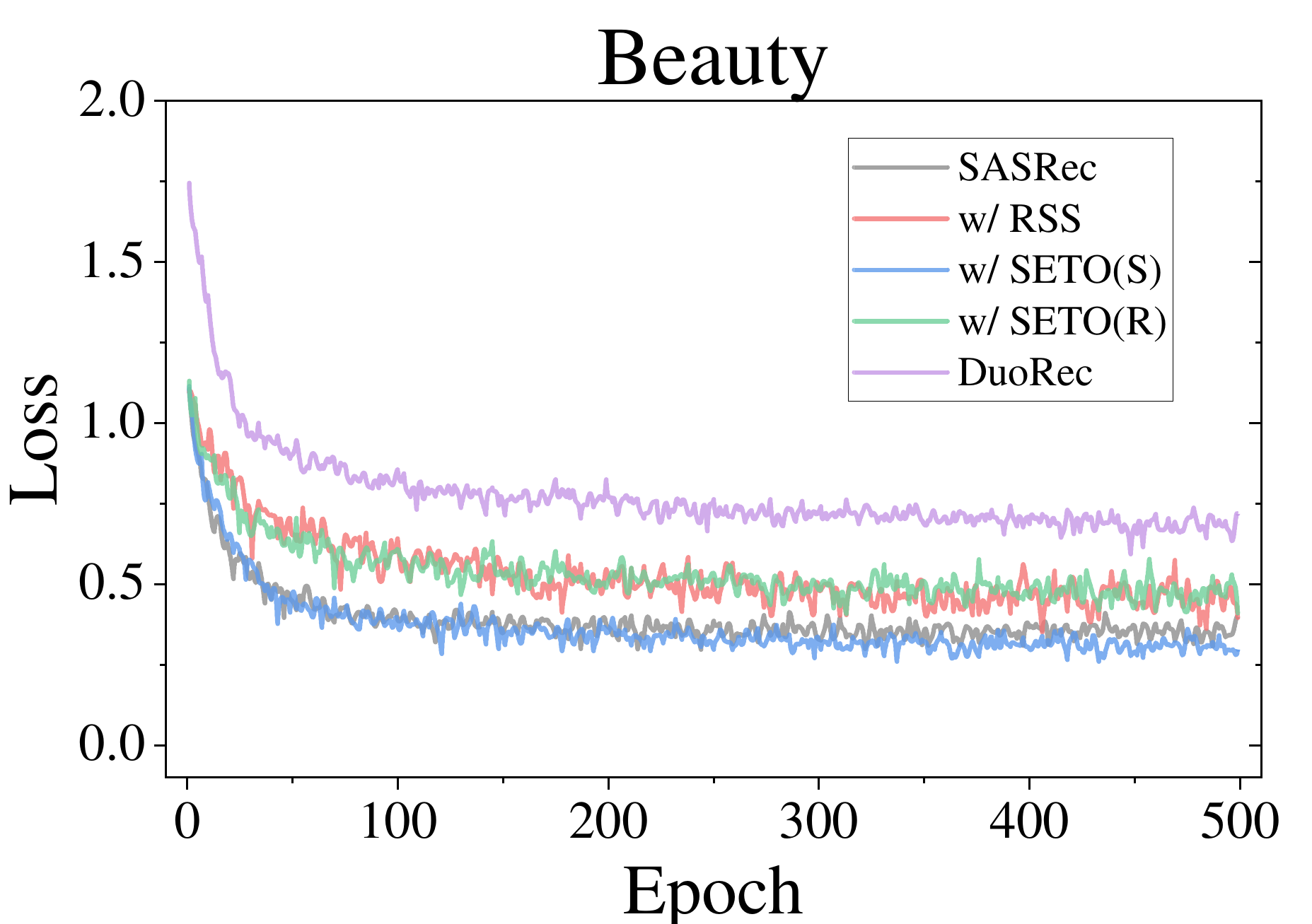}
    \caption{Training loss curve of different methods on different datesets.}
    \label{fig:loss}
\end{figure*}
\begin{figure*}[t]
    \centering
    \includegraphics[width=0.33\linewidth]{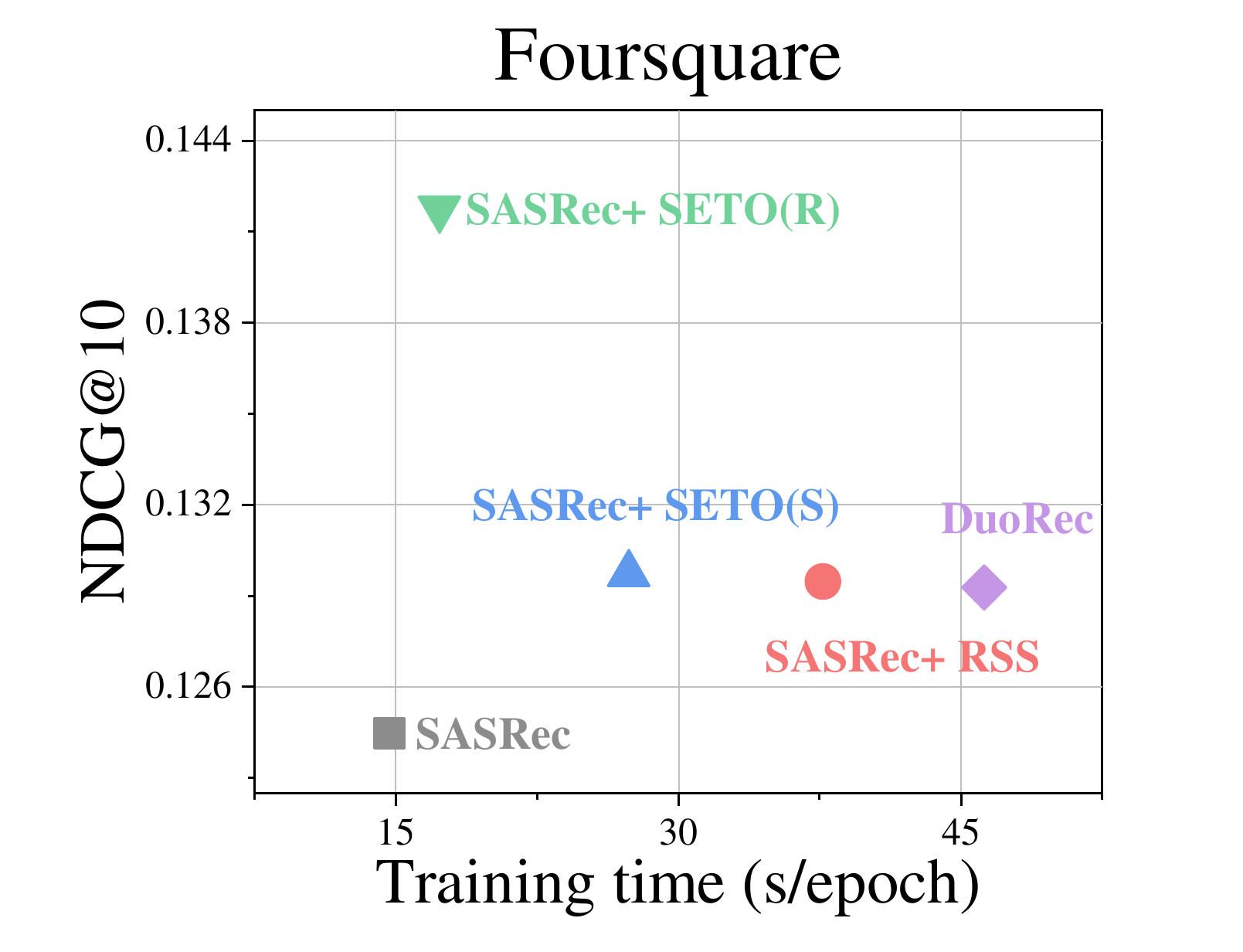}
    \includegraphics[width=0.33\linewidth]{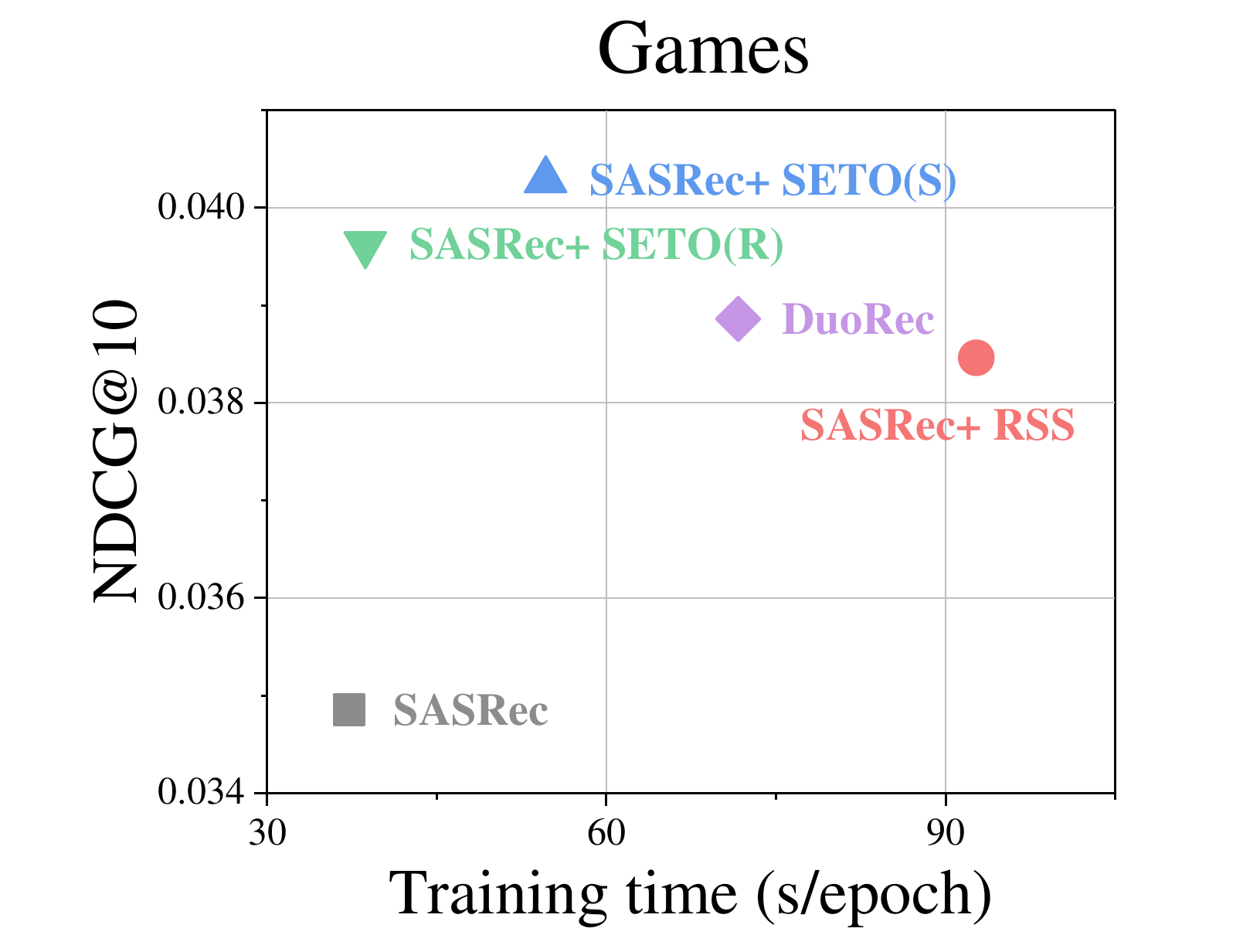}
    \includegraphics[width=0.33\linewidth]{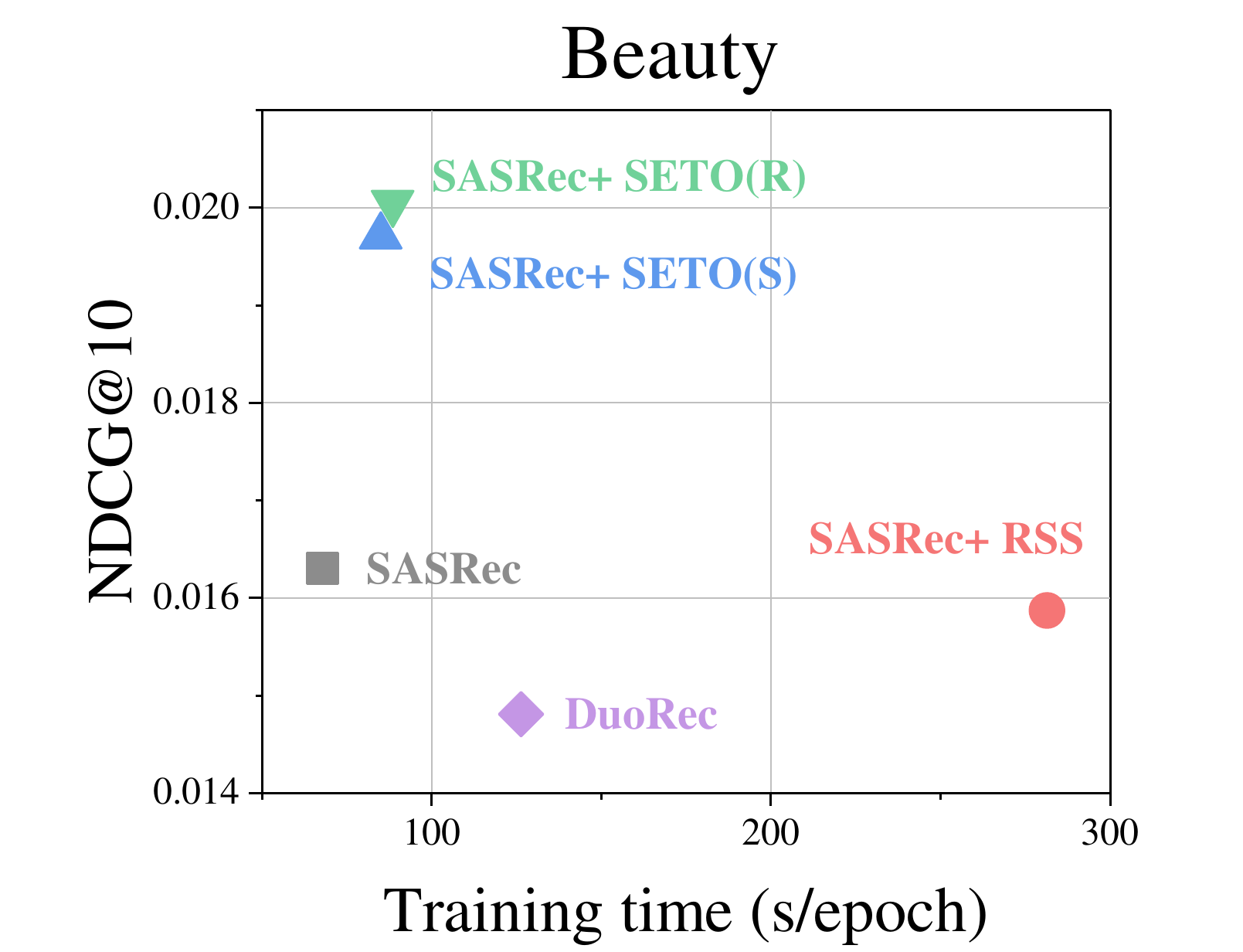}
    \caption{Training time (seconds per epoch) and recommendation performance of different methods on different datesets.}
    \label{fig:time}
\end{figure*}

\section{SETO vs. Traditional Enhancement}
\label{sec:vs}
Traditional data enhancement strategies are widely utilized, while our SETO distinguishes itself from the traditional methods and is novel from the following three perspectives:
\begin{itemize}[leftmargin=1.2em]
    \item In terms of the \textbf{enhancement objects}, existing methods only perform enhancement on the original sequences, ignoring the existence of a two-step transformation space (i) from input samples to preferences, and (ii) from preferences to output samples, in model training. We utilize the uncertainty in the transformation space to implement different randomized enhancement operations \textbf{on the subsequences}, which can obtain samples that cannot be obtained only on an original sequence.
    \item In terms of the \textbf{enhancement moment and duration}, instead of augmenting data before training as in previous works, we synchronize the operations with the training of the model, and do \textbf{temporary and dynamic augmentation} of the input and target subsequences in each iteration. It simulates the uncertainty that exists when converting high-dimensional preferences to low-dimensional data during training, and the temporary and dynamic operations alleviate the noise generated during enhancement to a certain extent.
    \item In terms of the \textbf{enhancement strategies}, the two operations $Swap$ and $Removal$ are different from the traditional operations. Moreover, we combine random selection and probabilistic selection under certain rationality constraints to realize the dynamics mentioned above.
\end{itemize}

\section{Algorithm of SETO}
 \label{sec:algorithm}
\begin{algorithm}
    \caption{Algorithm of our SETO to reconstruct samples on subsequences}
    \begin{algorithmic}[1] %每行显示行号
        \Require $S_u$: an interaction sequence of a user; $maxlen$: the maximum length of sequence for model training; $\alpha$: the hyperparameter of probability function in $Swap$; $scope$: the hyperparameter of $Swap$; $\rho$: the hyperparameter of $Removal$
        \Ensure $S_{input}$: the input subsequence in model training; $S_{target}$: the target subsequence in model training
        \Function {Sampler}{$s_{u}, maxlen, \alpha, scope, \rho$}
            \State \text{\# Sequential causal partitioning}
            \State $S_{input} \gets S_u[:-1]$
            \State $S_{target} \gets S_u[1:]$

            \State \text{\# Two augmentation operations on subsequences }
            % \State \text{\# with  Rationality Constraints}
            \If {operation== $Swap$}
            \State $S_{input} \gets $  \Call{Swap}{$s_{input},\alpha, scope$}
            \State $S_{target} \gets $  \Call{Swap}{$s_{target},\alpha, scope$}
            \ElsIf {operation== $Removal$}
            \State $S_{input} \gets $  \Call{Removal}{$s_{input},\rho$}
            \State $S_{target} \gets $  \Call{Removal}{$s_{target},\rho$}
            \EndIf

            \State \text{\# Padding short subsequences}
            \State $S_{input} \gets$ \Call{Pad}{$S_{input},maxlen$}
            \State $S_{target} \gets$ \Call{Pad}{$S_{target},maxlen$}

            \Return $S_{input},S_{target}$
        \EndFunction
    \end{algorithmic}
\end{algorithm}
\begin{algorithm}
    \caption{Algorithm of $Swap$ and $Removal$}
    \begin{algorithmic}[1] %每行显示行号
        \Require $S$: a sequence that needs to be enhanced; $f$: the probability function;  $\alpha$: the hyperparameter of probability functions in $Swap$; $scope$: the hyperparameter of $Swap$; $\rho$: the hyperparameter of $Removal$
        \Ensure $S'$: an enhanced sequence
        \Function {Swap}{$S, \alpha, scope$}
            \State $n \gets length(S)$
            \State $swapScope \gets max(int(n \times scope),1)$
            \State $indexA \gets random.choice(range(0\cdots n-1))$
            \State $Prob \gets f(swapScope,\alpha)$
            \State $indexB \gets random.choice(range(0 \cdots swapScope-1),Prob)$
            \State $indexB \gets indexA + random.choice([-1,1])\times indexB$
            \If {$indexB \in range(0 \cdots n-1)$}
            \State $S'=swap(S,indexA,indexB)$
            \EndIf
            
        \Return $S'$
        \EndFunction
        
        \Function {Removal}{$S, \rho$}
            \State $n \gets length(S)$
            \State $RemovalScope \gets random.choice(range(0\cdots n \times \rho))]$
            \State $S' \gets S$
            \For {each $i \in range(0\cdots RemovalScope)$}
                \State $n \gets length(S')$
                \State $index \gets random.choice(range(0\cdots n-1))$
                \State $S' \gets remove(S',index)$
            \EndFor
            
        \Return $S'$
        \EndFunction
        
    \end{algorithmic}
\end{algorithm}

\section{Backbone models}
 \label{sec:backbone}
We apply our SETO to various representative and state-of-the-art backbone models, including six single-domain sequential models and two cross-domain sequential models.
% Then we introduce a closely related work which is model-agnostic and without any extra context.

\begin{itemize}[leftmargin=1.2em]
    \item \textbf{Caser}~\cite{Caser} takes an ``image'' view of the relationships between items in user sequence and utilizes convolutional neural networks (CNNs) to extract the corresponding features.
    \item \textbf{SR-GNN}~\cite{SR-GNN} conveys and aggregates information from nearby nodes with graphical convolutional neural networks (GNNs) for session-based recommendation scenarios where the contained sequences are short.
    \item \textbf{SASRec}~\cite{SASRec} uses the self-attention mechanism to learn sequential relationships between items in a sequence, and has been experimented with as the representative model in several recent studies~\cite{PinnerFormer,meta,zhou2024contrastive}. 
    \item \textbf{FISSA}~\cite{FISSA} acts as an extension of SASRec to take the information of candidate items into account in the model. Note that it is more effective in some candidate sets.
    \item \textbf{FMLP-Rec}~\cite{FMLPRec} combines filtering algorithms with a full MLP architecture designed to adaptively filter the included noise information to help the model learn better.
    \item \textbf{DiffuRec}~\cite{DiffuRec} applies the diffusion model to the sequential recommendation, generating sequential item distribution representation and injecting uncertainty by corrupting the target item embedding into a Gaussian distribution via noise adding.
    \item \textbf{MGCL}~\cite{MGCL} applies contrastive learning on multi-view graphs to learn the static and dynamic collaborative filtering information from different domains.
    \item \textbf{TJAPL}~\cite{TJAPL} combines transfer learning and attentive preference learning to transfer information from the auxiliary domain to the target domain to solve sparse data issues.
\end{itemize}

\section{Experimental Details}
 \label{sec:details}
We follow their own code of the backbone models mentioned above but only modify the part of training sample construction. 
For each dataset, we adopt a leave-one-out evaluation methodology, where the second last item is used as validation and the last item is held for test.
Since FISSA is an extension of SASRec~\cite{SASRec}, it can be reduced to SASRec by adjusting the parameters. 
While selecting the internal parameters in the model, we only adjust the number of blocks at \{1,2,3\} because of its notable effect on performance. 
For DiffuRec~\cite{DiffuRec}, we follow the original setting of its code, where the dropout rate of turning off neurons in the Transformer block and item embedding is 0.1 and 0.3, and the number of Transformer heads is 4.
We find that the hidden dimensions must be a multiple of the number of heads in the code\footnote{{\href{https://github.com/WHUIR/DiffuRec}{ https://github.com/WHUIR/DiffuRec}}} of DiffuRec, so we set it to 52 and other methods keep it to 50.
% The dropout rate of turning off neurons in the Transformer block and item embedding is 0.1 and 0.3 for all datasets.
For the single-domain sequential recommendation methods, an early stopping strategy is used to evaluate the performance after every 50 epochs~\cite{FISSA}. In this strategy, the process stops if no improvement occurs in two consecutive evaluations. 
But for cross-domain models, due to the limitation of the running time of graph-based methods (i.e., MGCL~\cite{MGCL}) on the full candidate set, we follow \cite{MGCL} to use the partial candidate set, which contains randomly sampled 100 items that the user has not interacted with.
Note that our SETO\footnote{{\href{https://anonymous.4open.science/r/SETO-code-A026/}{The core source code: https://anonymous.4open.science/r/SETO-code-A026/}}} only temporarily handles the training sample part of each iteration and has nothing to do with the code of the model.

\section{Training Efficiency}
 \label{sec:efficiency}
 With the effective performance, we test the training efficiency of different methods (i.e., SASRec~\cite{SASRec}, RSS~\cite{RSS},DuoRec~\cite{DuoRec}, and our SETO).
 RSS is a model-agnostic approach that is similar to our SETO, while DuoRec is a model-dependent one based on contrastive learning.
The loss convergences are illustrated in Figure~\ref{fig:loss}.
 We measure the time consumed to train 100 epochs of each method and calculate the average time per epoch, and then combine the results with the recommended performance (taking NDCG@10 as an example) of different methods, as shown in Figure~\ref{fig:time}.
All experiments are conducted on an NVIDIA Tesla V100-PCIE-32GB GPU and Intel(R) Xeon(R) Gold 6240 CPU.

Figure~\ref{fig:loss} shows that the convergence speeds of these methods are close, which verifies that our SETO helps SASRec to get better performance without sacrificing the training convergence speed of the model.
In Figure~\ref{fig:time}, we observe that our SETO introduces only a slight increase in runtime to the original model while achieving significant improvements. 
Since RSS calculates the probability of an entire sequence, it takes more time on datasets with longer sequences, e.g., Games and Beauty.
Additionally, regarding the different variants of our SETO (i.e., SETO(S) and SETO(R)), the runtime of SETO(S) may exceed that of SETO(R) due to the additional computation required for the probability function.

\vspace{4mm}
\section{Discussion about the Amazon Dataset}
\label{sec:Amazon}
The Amazon\footnote{\href{url}{https://cseweb.ucsd.edu/\url{~}jmcauley/datasets.html\#amazon\_reviews}}
dataset is highly regarded and extensively utilized, and it stands as one of the few available publicly datasets for cross-domain and multi-domain sequential recommender systems~\cite{CDSRsurvey}.
A study~\cite{flawsOnAmazon} points out that the Amazon dataset has certain flaws because it contains review sequences rather than consumption sequences.
We agree that there are certain limitations, but believe that the Amazon datasets still have some significance in study of sequential recommendation methods.
There are some explanations as follows.

 Firstly, although a rating or review sequence is different from a purchase or consumption sequence, it is still a sequence. 
A recommender system can predict the items that a user will rate or review next, which is similar to predicting the items that a user will purchase or click next. 
In real life, many kinds of users' behaviors have a certain sequential nature, and studying those in different kinds of scenarios (not only in consumption scenarios) is a important step for sequential recommendation.

Secondly, about the problem of event conflict caused by temporal resolution~\cite{flawsOnAmazon}, it does make the task of modeling and learning sequential preferences more difficult.
However, it is a real kind of sequential recommendation scenario, which is worth studying. 
Moreover, it is worth mentioning that our augmentation strategy that utilizes the transformation space in subsequences is able to cope with this difficulty.

Thirdly, on a given dataset, the disparity in performance between models with and without sequential modeling is very much related to what kind of backbone model is used and how to improve the performance based on it.
It is also illustrated by some previous works~\cite{SASRec,FISSA,BERT4Rec,S3Rec,FMLPRec}, i.e., the gap between models without sequence modeling and models with sequence modeling is very significant on some Amazon datasets.

Moreover, we also apply other non-Amazon datasets (i.e., Foursquare and a large-scale industry dataset) to show the effectiveness of our SETO in sequential recommendation.
The Foursquare\footnote{\href{url}{https://archive.org/details/201309\_foursquare\_dataset\_umn}} dataset (which consists of check-in records based on locations and is often used to predict a user's travel order patterns) and the large-scale industry dataset (which consists of users' listening logs from a popular music streaming platform) that we use in the experiments in the paper, have a sequential nature and our method works very well on them.

\end{document}